\documentclass[useAMS,usenatbib,a4paper]{mn2e}
\usepackage{graphics}
\usepackage{graphicx}
\usepackage{epsfig}
\usepackage{amssymb,amsmath}
\usepackage{times}
\usepackage{natbib}
\usepackage{subfigure}

\title[Ram pressure in a galaxy formation model]{Ram pressure
  stripping in a galaxy formation model. I. A novel numerical approach}

\author[T. E. Tecce et al.]{Tom\'as
  E. Tecce$^{1,4}$\thanks{E-mail: tomas@iafe.uba.ar}, Sof\'ia
  A. Cora$^{2,4}$, Patricia B. Tissera$^{1,4}$, Mario
  G. Abadi$^{3,4}$ and \newauthor Claudia del P. Lagos$^{5,6}$\\
  ~\\
$^{1}$Instituto de Astronom\'ia y F\'isica del Espacio, C.C. 67 Suc. 28, 
  C1428ZAA Ciudad de Buenos Aires, Argentina\\
$^{2}$Facultad de Ciencias Astron\'omicas y Geof\'isicas, Universidad 
  Nacional de La Plata, and Instituto de Astrof\'isica de La Plata (CCT La
  Plata, CONICET,\\UNLP), Observatorio Astron\'omico, Paseo del Bosque
  S/N, B1900FWA La Plata, Argentina\\
$^{3}$Instituto de Astronom\'ia Te\'orica y Experimental and
  Observatorio Astron\'omico, Universidad Nacional de C\'ordoba,
  Laprida 854, X5000BGR C\'ordoba,\\~Argentina\\
$^{4}$Consejo Nacional de Investigaciones Cient\'ificas y T\'ecnicas, 
  Rivadavia 1917, C1033AAJ Ciudad de Buenos Aires, Argentina\\
$^{5}$Department of Physics, Institute for Computational Cosmology,
  University of Durham, South Road, Durham DH1 3LE\\
$^{6}$Departamento de Astronom\'ia y Astrof\'isica, Pontificia
  Universidad Cat\'olica de Chile, Av. Vicu\~na Mackenna 4860,
  Santiago, Chile}

\def \rvir{$r_\text{vir}$}
\def \mvir{$M_\text{vir}$}
\def \vvir{$V_\text{vir}$}

\def \msun{$M_\odot$}
\def \sag{{\scriptsize SAG}}
\def \sagrp{{\scriptsize SAGRP}}
\def \sagrpa{{\scriptsize SAGRP-A}}
\def \gadget {{\scriptsize GADGET-2}}
\def \fof {{\scriptsize FOF}}
\def \subfind {{\scriptsize SUBFIND}}

\begin{document}

\date{Accepted 2010 June 28. Received 2010 May 28; in original form
  2010 March 19}

\pagerange{\pageref{firstpage}--\pageref{lastpage}} \pubyear{2010}

\maketitle

\label{firstpage}

\begin{abstract}
We develop a new numerical approach to describe the action of ram pressure
stripping (RPS) within a semi-analytic model of galaxy formation and
evolution which works in combination with non-radiative hydrodynamical
simulations of galaxy clusters. The new feature in our method is the use of
the gas particles to obtain the kinematical and thermodynamical properties
of the intragroup and intracluster medium (ICM). This allows a
self-consistent estimation of the RPS experienced by satellite galaxies. We
find that the ram pressure in the central regions of clusters increases
approximately one order of magnitude between $z$~=~1 and~0, consistent with
the increase in the density of the ICM. The mean ram pressure experienced
by galaxies within the virial radius increases with decreasing redshift. In
clusters with virial masses \mvir~$\simeq$~10$^{15}\,h^{-1}$~\msun, over
50 per cent of satellite galaxies have experienced ram
pressures~$\sim$~10$^{-11}\,h^2$~dyn~cm$^{-2}$ or higher for $z
\lesssim$~0.5. In smaller clusters
(\mvir~$\simeq$~10$^{14}\,h^{-1}$~\msun) the mean ram pressures are
approximately one order of magnitude lower at all redshifts. RPS has a
strong effect on the cold gas content of galaxies for all cluster
masses. At $z$~=~0, over 70 per cent of satellite galaxies within the
virial radius are completely depleted of cold gas. For the more massive
clusters the fraction of depleted galaxies is already established at
$z \simeq$~1, whereas for the smaller clusters this fraction increases
appreciably between $z$~=~1 and~0. This indicates that the rate at
which the cold gas is stripped depends on the virial mass of the host
cluster. Compared to our new approach, the use of an analytic profile
to describe the ICM~results in an overestimation of the ram pressure
larger than 50 per cent for $z>$~0.5. 
\end{abstract}


\section[]{Introduction}
Comparison between galaxies in clusters and in the field reveals remarkable
differences in their physical properties, which suggest the important
influence of environment on galaxy formation and evolution. Large galaxy
surveys have shown that the colour distribution of the galaxy population is
clearly bimodal \citep{strateva2001,baldry2004}. One peak is formed by red,
non star-forming galaxies, while the other is populated by blue galaxies
with active star formation. In the local Universe, the fraction of red
sequence galaxies at a given stellar mass depends on environment
\citep{balogh2004,baldry2006}, with galaxy clusters having larger fractions
of red, early-type galaxies than the field. Environment also influences the
star formation rates (SFRs) of galaxies, as these are strongly suppressed
in dense environments for galaxies over a large range of stellar masses
\citep[e.g.][]{kauff2004}.

Another quantity which shows dependence on environment is galaxy
morphology. Spiral galaxies tend to be rarer in the central regions of
clusters than early-type galaxies \citep*{dressler80,whitmore93}, and
cluster spiral galaxies differ from those in the field in several
characteristics that can be correlated with environment \citep[for an
  extensive review see][]{boselli2006}. In clusters, spiral galaxies are
not only redder but also more H{\small I} deficient than similar galaxies
in the field, with this deficiency increasing for galaxies closer to the
cluster centre (\citealt*{haynes84}; \citealt{solanes2001,hc2009}). In
field spirals the gaseous discs typically extend beyond the optical radius,
but the opposite trend is seen in clusters, where a significant fraction of
spirals have truncated gaseous discs
\citep*[e.g.][]{koopk2004,koop2006,khc2006}.

A likely scenario is that galaxies are transformed from blue star-forming
systems into the red and passive population, but it is still unclear which
physical processes are the key ones responsible for this transformation, and
what their relative importance may be. The observed differences between
cluster and field galaxies outlined above suggest that the transformation
could be the result of processes that remove gas from galaxies, suppressing
their star formation, and/or alter the galaxy morphology, both effects
happening preferentially in high-density environments. Galaxy-galaxy
interactions and mergers are among these possible mechanisms as suggested
by some theoretical \citep{moore96,perez2006a,perez06b} and observational
\citep[e.g.][]{ellison2008,ellison2010,perez2009}
results. \citet{moore96,moore99} proposed that the evolution of
galaxies in groups and clusters is governed by the combined action of
multiple close encounters between galaxies and the tidal interaction
with the group potential, a process that they call `galaxy
harassment'. In clusters, however, the relative velocities of
galaxies are higher than in galaxy groups and, consequently, the
interaction times are shorter \citep{mihos2004}, so this process is
expected to be more significant in less dense environments.

Another process that might be relevant is the removal of the hot diffuse
gas halo of a galaxy after its infall into a group or cluster. This
has been called `strangulation' or `starvation'
\citep*{larson80,balogh2000,kawata2008}. The removed gas becomes part of the
overall intragroup or intracluster medium (ICM), the affected galaxy cannot
accrete any more gas via cooling flows, and with a moderate SFR it will
consume all of its cold gas within a few Gyr, ending its star formation and
becoming gradually redder as its stellar population ages. Strangulation is
a standard ingredient in most semi-analytic models of galaxy formation,
which successfully reproduce observed global properties of galaxies such as
luminosity functions, colour distributions and mass-metallicity relations
(e.g. \citealt*{baugh96}; \citealt{springel01,bower06,croton06};
\citealt*[][hereafter LCP08]{lcp08}). As commonly
implemented, as soon as a galaxy becomes a satellite, its hot gas halo is
assumed to be shock heated to the virial temperature of the group and
immediately removed from the galaxy. This effect proceeds in the same
fashion in groups of all masses.

Ram pressure stripping (RPS) of the cold gas in galactic discs could also
play an important role. Galaxies in clusters move through the hot diffuse
ICM gas at velocities that could be close to supersonic \citep{fd2006}, and
so will experience considerable ram pressure (RP). \citet[][hereafter
  GG72]{gg72} proposed that when the RP exceeds the gravitational restoring
force of the galaxy, its cold gas will be pushed out. In recent years, RPS
has been extensively studied using hydrodynamical simulations of individual
galaxies (e.g. \citealt*{abadi99,quilis2000,marcolini2003};
\citealt{roediger2006,roediger2007,kronberger2008}) which suggest
that the GG72 estimate is a fairly good approximation in most situations,
and that the time-scale for gas removal is $\sim$~10 -- 100~Myr. 
RPS acts only on the gaseous components of the galaxy, so its
characteristic signature is the presence of truncated gas discs while the
stellar discs remain unaltered. Indeed, there are several observational
candidates for RPS
(e.g. \citealt{crowl2005,bbc2006,cortese2007};
\citealt*{sun2007,vollmer2009}). RPS also seems to be relevant for
dwarf galaxies in less dense environments, such as galaxy groups
(\citealt{mc2007}; \citealt*{mastropietro2008}). A recent review of
both simulations and observations of RPS can be found in
\citet{roediger2009}.

All the physical processes described above are not mutually exclusive. In a
hierarchical universe, many galaxies in clusters were previously members of
smaller systems where a combination of strangulation, harassment and/or
RPS could start to suppress star formation \citep[this has been called
  `galaxy pre-processing'; see
  e.g.][]{fujita2004,mihos2004,cortese2006,perez2009}.

Semi-analytic models of galaxy formation include galaxy mergers and
strangulation as standard elements, but the effect of RPS has been included
in such models only in few cases \citep{on2003,lanzoni2005} where no
significant influence of RPS was found on the analysed galaxy
properties. These previous works use dark matter (DM)-only simulations to
generate the merging history trees of DM haloes, which are then used
by the semi-analytic model to generate the galaxy population, and the
properties of the ICM and velocity distributions of galaxies are
modeled using analytical approximations.

A more recent work by \citet[][hereafter BDL08]{bdl2008} studies the
distribution and history of the RP experienced by galaxies in clusters,
combining a semi-analytic model with the Millennium Simulation
\citep{millennium}; they use the properties of the DM particles to
track the positions and velocities of simulated galaxies. Again, since
the Millennium is a DM-only simulation, BDL08 have to resort to
analytic models for the ICM properties, assuming the hypothesis of
hydrostatic equilibrium for the gas within a DM halo described by the
\citet*[][hereafter NFW]{nfw97} profile. The latter is a good
approximation to the DM profiles if the haloes are in dynamical
equilibrium; this is likely not the case for high-redshift haloes,
where the bulk of the star formation activity takes place
(e.g. \citealt{madau96}; \citealt*{ciardi2003};
\citealt{hopkins2007}). Since the dynamics of the ICM may play an
important role \citep*{sunyaev2003} and certainly the hypothesis of
hydrostatic equilibrium may not hold for all haloes, our aim is to
develop an improved model for RPS which takes into account the ICM
dynamics through simulations that include gas physics.

We model RPS by adopting the criterion proposed by GG72 and by
implementing this process in the semi-analytic model of galaxy formation
and evolution Semi-Analytic Galaxies~(\sag; LCP08), which
is combined with hydrodynamical cosmological simulations of galaxy
clusters \citep{dolag05}. The novel feature of our implementation is
the fact that the thermodynamical and kinematical properties of the
ICM, which are involved in the estimations of the RP experienced by
each galaxy, are provided by the underlying simulations. In this
paper, the first of a series, we focus on the study of the
distributions of RP experienced by satellite galaxies in clusters of
different masses, the dependence of RP with clustercentric distance,
and how these quantities evolve with redshift. We compare our results
with those obtained by assuming NFW profiles for the density
distribution of the ICM. A forthcoming paper will deal with the
influence of RPS on galaxy properties such as luminosities, colours
and star formation histories.

This paper is organised as follows. In Section~\ref{sec:model}, we briefly
describe the semi-analytic model \sag~used in this work and give details of
the simulations used. Section~\ref{sec:rps} contains a detailed description
of the way in which RP is estimated from the hydrodynamical simulations,
and the implementation of the RPS process in the semi-analytic
model. Section~\ref{sec:results} presents the distributions of RP that we
obtain from our cluster simulations and compare them to the results
obtained by adopting a NFW profile for the ICM under the hypothesis of
hydrostatic equilibrium. We then analyse the effects of RPS on the gas
content of galaxies. Finally, in Section~\ref{sec:conclu} we summarize our
main findings.


\section[]{Numerical simulations and the semi-analytic model}
\label{sec:model}
To develop our RPS model we use a hybrid numerical approach which combines
cosmological non-radiative $N$-body/smoothed particle hydrodynamics (SPH)
simulations of galaxy clusters with a semi-analytic model of galaxy
formation and evolution. In the cosmological simulation, the DM haloes and
their substructures are identified and followed in time, building up
detailed merger trees which are used by the semi-analytic code to generate
the galaxy population. The main advantage of this hybrid procedure is that
the use of semi-analytic codes allows the exploration of a larger dynamical
range than fully self-consistent simulations, at a fraction of the
computational cost.

In the present work we implement the RPS process in the semi-analytic code
developed by LCP08. We take the kinematical and thermodynamical properties
of the hot diffuse gas which permeates DM haloes directly from the
underlying SPH simulation, thus avoiding the use of analytical
approximations. In this section, we briefly describe the simulations and
the semi-analytic model on to which we graft our RPS scheme described in
detail in Section~\ref{sec:rps}.

\subsection[]{Non-radiative $N$-body/SPH simulations of galaxy clusters}
\label{sec:sims}
In this study, we use the high-resolution hydrodynamic non-radiative
simulations of galaxy clusters by \citet{dolag05}. The simulated clusters
were originally extracted from a DM-only simulation with a box size of
479~$h^{-1}$~Mpc of a flat $\Lambda$CDM model with $\Omega_{\rm m}$~=~0.3,
$\Omega_{\Lambda}$~=~0.7, $H_0$~=~100~$h^{-1}$~km~s$^{-1}$~Mpc$^{-1}$ where
$h$~=~0.7, a baryon density $\Omega_b$~=~0.039 and a normalization of the
power spectrum $\sigma_8$~=~0.9 \citep*{yoshida2001}. The Lagrangian
regions surrounding the selected clusters have been resimulated at
higher mass and force resolution using the `zoomed initial conditions'
technique \citep*{tormen97}. Gas was introduced in the high-resolution
region by splitting each parent particle into a gas and a DM
particle. The mass resolution is the same for all simulations, being
$m_\text{DM}$~=~1.13~$\times$~10$^9 \,h^{-1}$~\msun~for DM particles
and $m_\text{gas}$~=~1.69~$\times$~10$^8 \,h^{-1}$~\msun~for gas
particles. However, the identification of DM haloes was based only on
the DM particles, with their mass increased to its original value. 

The simulations have been carried out using \gadget, a parallel Tree-SPH
code with fully adaptive time-stepping and explicit conservation of energy
and entropy \citep{springel05}. For the force resolution, the gravitational
softening is fixed at $\epsilon$~=~5~$h^{-1}$~kpc in physical units at $z
\leq$~5, and for higher redshifts it shifts to $\epsilon$~=~30~$h^{-1}$~kpc
in comoving units. For each simulation, 92 snapshots were stored between
redshifts $z$~=~60 and $z$~=~0. The simulations considered here include
only non-radiative physics and the original formulation of artificial
viscosity within SPH; this is a reasonable approximation for
simulations of clusters, since the bulk of the halo gas has very long
cooling times. 

From these simulations, we have selected eight cluster-sized haloes
divided in two sets: five clusters with \mvir~$\simeq$~10$^{14} \, h^{-1}
M_\odot$ (hereafter, G14 clusters) and three clusters with
\mvir~$\simeq$~10$^{15} \, h^{-1} M_\odot$ (G15 clusters). These DM
haloes are first identified in the simulation outputs by means of a
friends-of-friends (\fof) algorithm \citep{davis85}. Subsequently,
the \subfind~algorithm \citep{springel01} is applied to the haloes
detected by \fof~in order to find self-bound DM substructures which we
call {\it subhaloes}. To build the trees, we extract from the simulations
all subhaloes consisting of 10 or more DM particles, since smaller ones are
usually dynamically unstable \citep{kauff99}. Additionally, SPH particles
in the simulations provide the density, spatial and velocity distributions
of the intergalactic and intracluster media.

\subsection[]{Semi-analytic model of galaxy formation}
\label{sec:sag}
The DM halo merger trees and the information from the SPH particles are
then used as input for the semi-analytic model of galaxy formation
\sag~(LCP08), based on the one described by \citet{cora06}, which includes
the effects of radiative cooling of hot gas, star formation, feedback from
supernovae explosions, chemical enrichment and galaxy
mergers. \sag~includes the effect of strangulation as described in the
Introduction. LCP08 updated it to include black hole (BH) growth and
feedback from active galactic nuclei (AGN). In addition to this major
improvement, \sag~allows starbursts in three different ways: both in major
and minor mergers, and when disk instabilities occur. The reader is
referred to \citet{cora06} and LCP08 for the full details of these
implementations. In the present work, this model is further modified to
include the effect of RPS on galaxies due to the hot intergalactic gas as
described in the next section.

The galaxy catalogue is built up by applying the semi-analytic model to the
detailed DM subhalo merger trees extracted from the hydrodynamical
simulations. Similarly to other semi-analytic models, in this `subhalo
scheme' arising from the identification of DM substructures within
\fof~groups, the largest subhalo in a \fof~group is assumed to host the
central galaxy of the group, located at the position of the most bound
particle of the subhalo. These galaxies are designated as {\it central} or
{\it type~0} galaxies, and each \fof~halo has only one. Central galaxies
of smaller subhaloes contained within the same \fof~group are
referred to as {\it halo} or {\it type~1} galaxies. The subhaloes of these
galaxies are still intact after falling into larger structures. There is a
third group of galaxies generated when two subhaloes merge and the galaxy
of the smaller one becomes a satellite of the remnant subhalo: these
galaxies are called {\it type~2} galaxies. A type~0 galaxy can have
satellites of types~1 and~2 and, furthermore, a type~1 galaxy may itself
have type~2 satellites. In the following, whenever we use the term
`satellite galaxies' we will be referring to both type~1 and~2
galaxies. We assume that type~2 galaxies merge with their
corresponding subhalo central galaxy on a dynamical friction
time-scale.

For each of these galaxies, \sag~provides information on the stellar mass,
cold disc gas, hot gas within DM haloes, BH mass, AGN activity, star
formation histories, magnitudes in several bands according to the stellar
population models by \citet{bc03} and chemical abundances of the different
baryonic components. Galaxy positions and velocities are traced by tracking
the position and velocity of the most bound particle of their host
DM~subhalo; for type~2 galaxies, we use the most bound particle identified
at the last time there was a substructure (as in BDL08). In a
forthcoming paper we will test how the effect of RPS depends on the
choice of alternative recipes for the different physical processes
such as the star formation scheme and the galaxy orbits.

\section{Ram pressure stripping of cold disc gas}
\label{sec:rps}
In this Section we describe our self-consistent implementation of RPS in
the semi-analytic code. Our RPS model is based on the simple criterion
proposed by \citet{gg72}: the cold gas of the galactic disc located beyond
galactocentric radius $R$ will be stripped away if the RP exerted by the
ambient medium on the galaxy exceeds the restoring force per unit area due
to the gravity of the disc:
\begin{equation}\label{eq:rps}
  \rho_\text{ICM}v^2 \geq 2\pi G
  \Sigma_\text{disc}(R)\Sigma_\text{cold}(R).
\end{equation}
Here $\rho_\text{ICM}$ is the ambient density of the ICM at the current
position of the galaxy, $v$ the velocity of the galaxy relative to the ICM
and $\Sigma_\text{disc}$, $\Sigma_\text{cold}$ are the surface densities of
the galactic disc (stars plus cold gas) and of the cold gas disc,
respectively. Although the description of RP phenomena is usually quoted in
the context of galaxy clusters, it is valid in general, and we apply this
criterion to the hot gas contained within all DM haloes at all redshifts. 

In order to determine the RPS, we need an estimation of each of the
parameters involved in condition~\eqref{eq:rps}, namely the properties of
the ICM on the left-hand side and the scalelength of the galactic discs
necessary to evaluate the right-hand side. Below, we explain the strategies
used to estimate these quantities within our hybrid model.

\begin{figure}
  \centering
  \subfigure {
    \epsfig{file=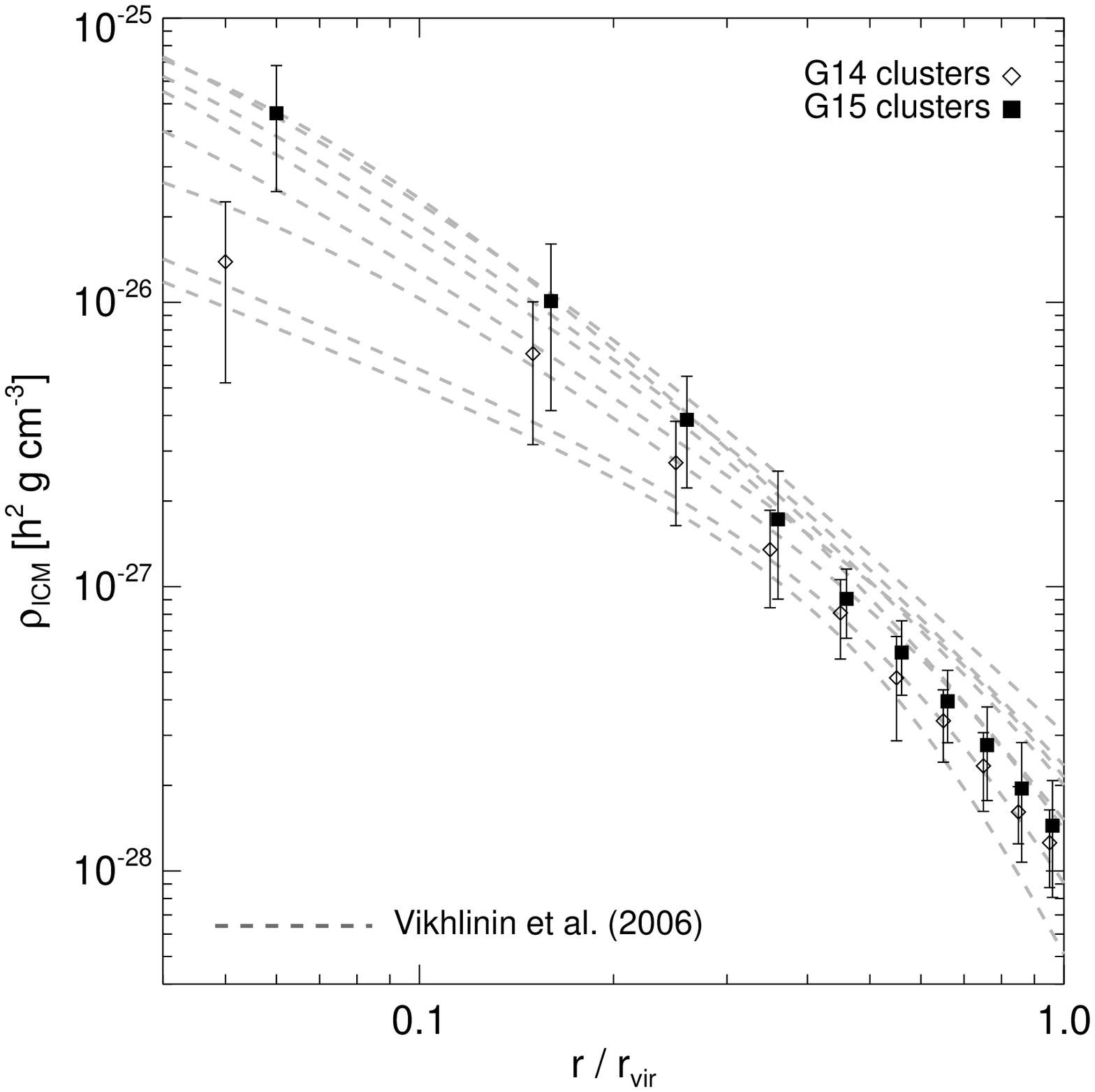, width=0.45\textwidth}
  }
  \subfigure {
    \epsfig{file=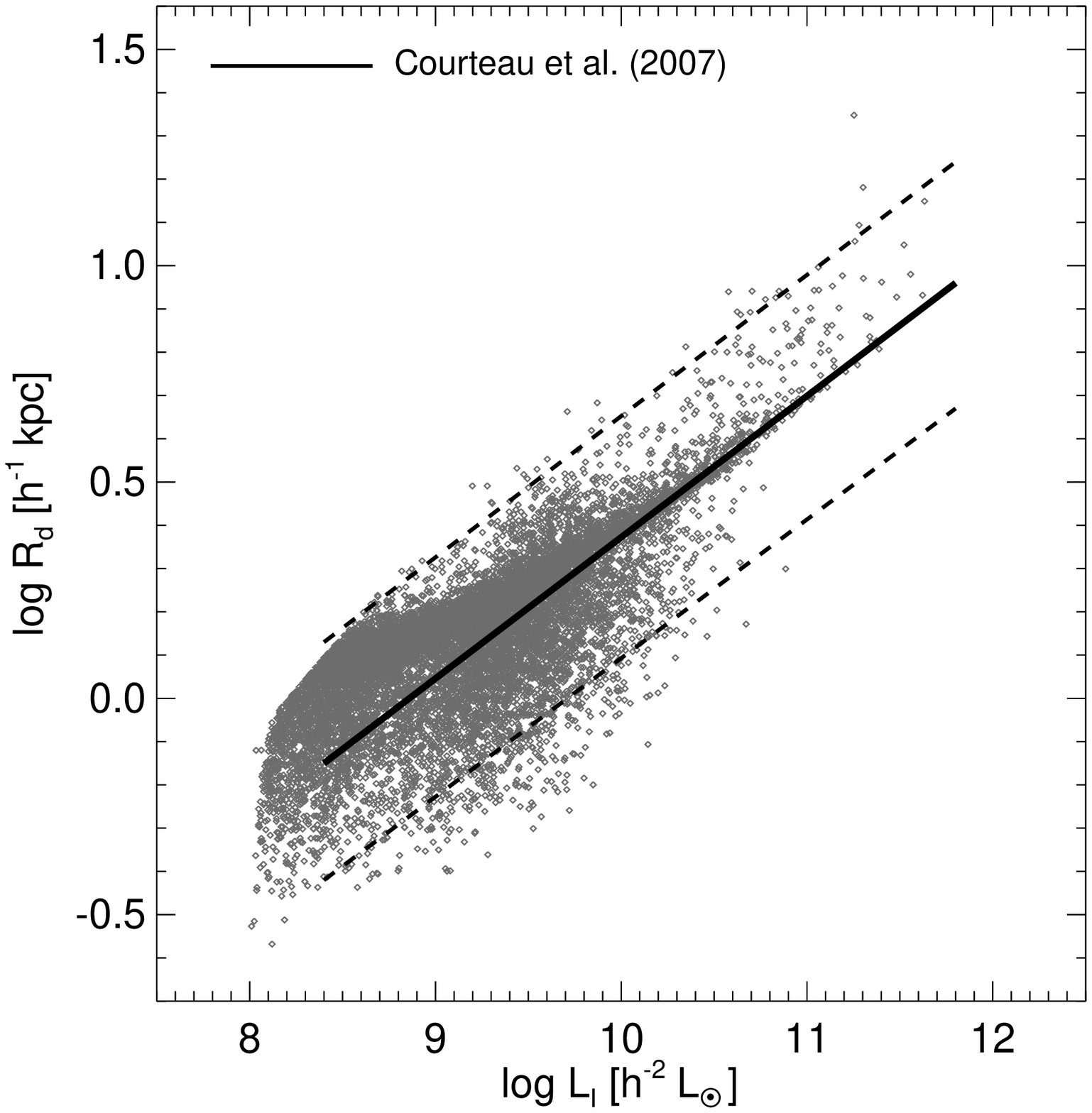, width=0.45\textwidth}
  }
  \caption{Top: mean ICM density profiles of the simulated clusters
  at $z$~=~0. Open diamonds and filled squares show values for the G14 and
  G15~clusters, respectively. Data points for the G15~clusters are shifted
  to the right for clarity. Dashed lines show the average density profiles
  determined by \citet{vik06} for a sample of nearby relaxed clusters. 
  Bottom: relation between disc scalelength $R_d$ and $I$-band
  luminosity for disc galaxies, for a run of the \sag~model that includes
  the RPS effect. We selected galaxies with $M_\text{baryon}
  \geq$~10$^8\,h^{-1}$~\msun, $M_\text{bulge}/M_\text{stellar} \leq$~0.95
  and with central surface brightness $\mu_0(B) <$~22 mag
  arcsec$^{-2}$. For this plot we take all the galaxies in one of the
  largest simulation boxes, including both cluster-sized haloes and smaller
  systems. Solid and dashed lines show the mean observed relation and
  2$\sigma$ scatter from \citet{courteau07}.}
  \label{fig:sagnew}
\end{figure}

\subsection{Determination of the ICM properties}
\label{sec:icmprop}
As already mentioned in the Introduction, the main difference between our
RPS implementation in a semi-analytic model and previous ones
\citep{on2003,lanzoni2005} is that we use the thermodynamical properties of
the ICM derived from the underlying hydrodynamical simulation. For all
satellite galaxies in the simulations, we determine the local ICM density
$\rho_\text{ICM}$ and velocity~$v$ relative to the ICM by searching for all
gas particles in a sphere of radius $r_\text{ngb} \propto$~\rvir~centred
on each galaxy, where \rvir~is the virial radius of the host DM subhalo of
the satellite. For type~2 galaxies, the value of \rvir~corresponds to the
last time the galaxy was identified as either type~0 or type~1. If less
than 32 particles are found, we take the closest 32~neighbours instead. 

Some of the particles selected by this procedure might be gravitationally
bound to the DM~subhalo. In such a case, ICM densities would be
overestimated, and the velocity of the galaxy relative to the mean ICM
would artificially be $v \sim$~0. Thus, these particles need to be
discarded since we are interested in estimating the properties of the
ambient medium which is responsible for the RPS. To do this, once all gas
neighbour particles have been found, the higher density ones are filtered
out by using an iterative procedure: we determine the median gas density
$\rho_m$ of the selected particles, discard all particles with $\rho >
f_\text{rm}\rho_m$ and repeat this procedure until the median density
converges. We find that choosing $r_\text{ngb}$~=~2.5~\rvir~and
$f_\text{rm}$~=~2, a smooth ICM density distribution can be recovered,
removing substructure in the gas without significantly affecting the median
density profile. The final step is then to determine the velocity $v$ of
the galaxy relative to the mean motion of the gas particles that remain
after filtering.

By calculating the ICM properties in this fashion, we obtain a
self-consistent method which does not introduce any additional free
parameter into the model and does not need to assume that the ICM gas is in
hydrostatic equilibrium. In fact, it automatically takes into account any
local variation of the density and/or the velocity field due to the
dynamics of the gas. The method is not overly sensitive to the values of
$f_\text{rm}$ and $r_\text{ngb}$ chosen; $r_\text{ngb}$ is a compromise
between the need to consider a radius large enough to obtain a fair sample
of the local environment, but not so large that the search spans a large
portion of the cluster volume. With $f_\text{rm} >$~3 the method becomes
ineffective at removing substructure, and choosing $f_\text{rm} \leq$~1 the
method underestimates the median ICM density when compared to observations.

The density profiles we obtain from this procedure, for the case of
the cluster-sized haloes, are in excellent agreement with those
determined from X-ray observations of galaxy clusters at $z \simeq$~0
(e.g. \citealt*{schindler99}; \citealt{vik06}). This can be seen in
the top panel of Fig.~\ref{fig:sagnew}, where we compare the mean
ICM~density profiles obtained for the simulated clusters with the
profiles obtained by \citet{vik06} for a sample of nearby relaxed
clusters, which span a temperature range 0.7~$-$~9~keV that comprises
the mass-weighted temperatures of the simulated clusters
\citep[see][]{dolag05}.

\subsection{Sizes of galactic discs}
\label{sec:sizes} 
In the semi-analytic code \sag, the gas acquired by a galaxy via cooling
flows is assumed to settle onto a thin exponential disc with surface
density $\Sigma (R) = \Sigma_0 \exp(-R/R_d)$, where $\Sigma_0$ is the
central surface density and $R_d$ the disc scalelength. If the outer disc
radius $R_\text{disc} \gg R_d$, then these quantities are related to the
total disc mass~$M_d$ through 
\begin{equation}\label{eq:sigma0}
 \Sigma_0 = \frac{M_d}{2\pi R_d^2}.
\end{equation}
To determine the disc scale length we follow the model developed by
\citet*[hereafter MMW]{mmw98}. Assuming a NFW profile to represent the DM
distribution, the MMW model allows the calculation of $R_d$ given the halo
virial mass \mvir, the mass fraction of baryons that settle on to the disc
$m_d$~=~$M_d/$\mvir, the halo concentration $c$ and the dimensionless halo
spin parameter $\lambda = J |E|^{1/2}G^{-1}M_\text{vir}^{-5/2}$, where $J$
and $E$ are the angular momentum and energy of the halo, respectively.

The MMW model assumes that the DM halo responds adiabatically to the slow
assembly of the disc, following the standard scheme of \citet{blumen86}.
The initial NFW mass profile $M(r_i)$ and the profile after contraction
$M_f(r)$ are then related by 
\begin{equation}\label{eq:mmwmass}
  M_f(r) = M_d(r) + M_b + M(r_i)(1-m_d-m_b),
\end{equation}
where $M_b$ is the total bulge mass (considering bulges as point masses,
for simplicity), $m_b$~=~$M_b/M_\text{vir}$, and $M_d(r)$ is the disc mass
within $r$ given by 
\begin{equation}\label{eq:discmass}
  M_d(r) = 2\pi \Sigma_0 R_d^2 
  \left[1 - \left(1+\frac{r}{R_d}\right)\exp(-r/R_d) \right].
\end{equation}
The implicit assumption is that the initial distribution of baryons
parallels that of the DM, and those that do not end up in the disc or in
the bulge remain distributed in the same way as the DM. These
considerations allow MMW to express the disc scalelength as
\begin{equation}\label{eq:rd}
  R_d = \frac{1}{\sqrt{2}}\left(\frac{j_d}{m_d}\right)\lambda \,r_\text{vir}
  f_c^{-1/2}f_R(\lambda,c,m_d,j_d),
\end{equation}
where $j_d$~=~$J_d/J$, with $J_d$ the total disc angular momentum, and
$f_c$, $f_R$ are factors given by
\begin{equation}
  f_c \simeq \frac{2}{3} + \left(\frac{c}{21.5}\right)^{0.7},
\end{equation}
\begin{equation}\label{eq:fr}
  f_R(\lambda,c,m_d,j_d) = 2\left[\int_0^\infty e^{-u}u^2
    \frac{V_c(R_du)}{V_\text{vir}} du\right]^{-1}
\end{equation}
(see MMW for the full details). If the specific angular momentum of the
disc material is the same as that of the parent halo, then the ratio
$j_d/m_d$ should be close to unity \citep{fall80}. Thus, we set $j_d = m_d$
and assume the angular momentum of the bulge to be negligible.

For a given set of values of \vvir, $c$, $\lambda$, $m_d$ and
$j_d$, equations \eqref{eq:mmwmass}~-~\eqref{eq:fr} must be solved by
iteration to yield $R_d$. However, including this iterative method in the
semi-analytic model is extremely time consuming. Most semi-analytic models
use a fitting formula for $R_d$ determined by MMW (their equation~32),
but this is valid only for the case of pure disc galaxies and with an
accuracy of 15 per cent quoted only for values of $m_d$ in the range
0.02~$< m_d <$~0.2. We find that galaxies in our simulations cover a
much broader range in $m_d$; besides, we also want to take into
account the presence of bulges and their gravitational effect on the
size of the final disc. Hence, we use the iterative procedure to
generate a set of look-up tables for each snapshot of the underlying
simulation and for a grid of values of $m_d$, $m_b$, $c$ and
$\lambda$. These tables are then used by the semi-analytic model to
find $R_d$ by interpolation. 

We assume that the process of disc formation takes place only for central
galaxies. When a galaxy becomes a satellite, the accretion of cold gas is
halted due to strangulation, so its disc cannot grow any further. We then
take $R_d$ to be frozen at the value it had at the last time the galaxy was
identified as a central. Additionally, if a galaxy suffers a major merger
or if the galactic disc becomes unstable, the remnant is a spheroid and so
$R_d$ is set equal to zero after such an event.

The values of $m_d$ and $m_b$ are calculated for each central galaxy from
the properties given by the semi-analytic code. The concentration of their
host DM haloes are determined as a function of \mvir~and redshift by
following the model by \citet{bullock2001}, with the parameters quoted by
\citet{wechsler2006}. The spin parameter $\lambda$ is determined for each
halo by computing its total energy and angular momentum. Studies based in
$N$-body simulations have found that the distribution of $\lambda$ for
DM~haloes can be well described by a lognormal function with mean
0.03~$\lesssim \lambda_0 \lesssim$~0.05 and dispersion $\sigma \simeq$~0.5
\citep[e.g.][]{warren92,bett2007,maccio07}. For the simulations used here
$\lambda_0$~=~0.042, but the distribution deviates from the lognormal for
high values of~$\lambda$. This happens because the mass resolution in
$N$-body simulations affects the determination of angular momenta,
independently of the method used to find DM structure, generating an
artificial high-$\lambda$ tail in the distribution \citep{bett2007}. The
value of the spin parameter correlates with the degree of equilibrium of
the DM halo, with haloes which are far from being virialized having large
values of $\lambda$. As a consequence, we use the simple criterion proposed
by \citet{bett2007} to filter out these anomalous values. For each halo, we
calculate an `instantaneous virial ratio'
\begin{equation}\label{qe}  
  Q \equiv |2T/U + 1|,
\end{equation}
and consider that haloes with values $Q < 1$ are in a quasi-equilibrium
state. This allows us to correct the distribution of $\lambda$ for mass
resolution effects. For central galaxies of anomalous haloes ($Q \geq 1$),
we set $R_d = \lambda_0$~\rvir$/\sqrt{2}$~(MMW, their equation 12).

This procedure generates a distribution of sizes of galactic discs which is
in good agreement with observations. The bottom panel of
Fig.~\ref{fig:sagnew} shows the relation between disc scalelength and
$I$-band luminosity, compared to the observed relation from
\citet{courteau07}. The data were extracted from a run of the
\sag~model which includes the RPS effect as described in
Sections~\ref{sec:icmprop} and \ref{sec:rpsmodel} (see also Section
\ref{sec:rpseff}). The plot includes all galaxies with
$M_\text{baryon} \geq$~10$^8\,h^{-1}$~\msun~within one of the largest
simulation boxes, comprising not only cluster-sized haloes but also
smaller systems. We select as spirals those galaxies with central
surface brightness $\mu_0(B) <$~22 mag arcsec$^{-2}$ to exclude low
surface brightness (LSB) systems \citep{impey2001}, and with
$r_\text{thresh} \equiv M_\text{bulge}/M_\text{stellar} \leq$~0.95
(following LCP08). Varying $r_\text{thresh}$ between 0.5 and 0.95 does not
result in any significant change in the distribution of morphological types
with galaxy stellar mass.

The agreement between the model and observations is very good both in terms
of the slope and the scatter of the relation. For a given luminosity,
galaxies with higher values of $\lambda$ (i.e. with higher angular
momentum) have larger discs. Once the artificially large values of
$\lambda$ have been corrected, very few galaxies ($<$~5 per cent) fall
in the region above the upper dashed line in Fig.~\ref{fig:sagnew},
and those are mostly LSB galaxies. Disc galaxies are not found in the
region below the lower dashed line (and below the mean for $L_I
>$~10$^{11}\,h^{-2} L_\odot$) because such discs become unstable.

\subsection[]{Modelling ram pressure stripping }
\label{sec:rpsmodel}
With the relevant properties of the ICM estimated as previously described
we can now construct our RPS model. The mass loss due to RPS is
calculated analytically as follows. For exponential discs of stars and
gas, it can be shown from condition~\eqref{eq:rps} that RPS will
remove from a galaxy all the gas beyond a stripping radius
$R_\text{str}$ given by   
\begin{equation}\label{eq:rstrip}
  R_\text{str} = - 0.5 R_d \ln \left[ {\frac{\rho_\text{ICM}
        \,v^2} {2\pi G \Sigma_{0,\text{disc}}
        \Sigma_\text{0,\text{cold}}}} \right].
\end{equation}
Here $\Sigma_{0,\text{cold}}$ and $\Sigma_{0,\text{disc}}$ are the central
surface densities of the cold gas disc and of the galactic disc. We assume,
for simplicity, that both disc components (stars and cold gas) have the
same scalelength. To avoid extremely large values of $R_\text{str}$, which
may lead to divergences in the calculations, we ignore the effect of RPS if
$R_\text{str}/R_d \geq$~20; in such cases the fraction of gas removed is
negligible. 

Our calculation of mass loss follows that of \citet{lanzoni2005}. The
previous study by \citet{on2003} used a much rougher estimation of
RPS: for each galaxy they calculated a mean gravitational restoring
force, and if the RP exceeded its value, all the cold gas was removed;
if the RP was smaller, no gas was lost. This is a crude assumption,
and we consider the \citet{lanzoni2005} approach a better
representation of the continuous gas removal caused by RPS. A
difference between our work and \citet{lanzoni2005} is that they
include a factor $\cos^2 i$ in the left-hand side of \eqref{eq:rps} to
account for the disc inclination, where $i$ is the angle between the
normal to the plane of the disc and the direction of motion of the
galaxy through the ICM. We have chosen to neglect this factor because
studies based on SPH simulations agree that the galaxy inclination
does not have a significant influence on the amount of removed gas;
strong RPs that strip a face-on galaxy also strip a galaxy moving
edge-on, although on a slightly longer time-scale \citep[see][and
  references therein]{roediger2009}. We have also chosen to ignore the
possibility of enhanced star formation as a result of RP compressing
the molecular gas of the galaxy
\citep{fujita99,bekki2003,kronberger2008,kapferer2009}, since
observational evidence is contradictory
\citep{khc2006,koop2006,ak2009} and theoretical work needs
confirmation due to the complex nature of star formation.

In \sag~the evolution of the galaxy population is followed by solving
differential equations describing the physical processes involved, such as
gas cooling, star formation and feedback, at small time-steps of
size~$\Delta T/N_\text{steps}$, where~$\Delta T$ is the time interval
between outputs of the underlying cluster simulation; we adopt
$N_\text{steps}$~=~50. Since the hydrodynamical properties of the ICM and
galaxy positions and velocities are given by the underlying simulation, the
effect of RPS is estimated for each galaxy only once per snapshot,
considering a single instantaneous stripping event and assuming that the
profile of the exponential disc remains unaltered for $R < R_\text{str}$.
The mass of gas removed by RPS is transferred to the hot gas component of
the central galaxy of its \fof~halo, thus becoming available to the central
galaxy for cooling and star formation in subsequent time-steps. 

The RPS effect eats away the gas disc from the outside in. After the first
stripping event, the remaining disc gas (if any is left) is assumed to form
an exponential disc with the same scalelength as before the stripping, but
sharply truncated at $R_\text{str}$. If in a subsequent snapshot the galaxy
experiences a higher RP, $R_\text{str}$ will then be smaller and the gas
lost to RPS will be the gas located between $R_\text{str}(t_n)$ and
$R_\text{str}(t_{n-1})$, where $t_n$ and $t_{n-1}$ denote the current and
previous simulation outputs, respectively. Since in our strangulation model
satellite galaxies have no infall of cooling gas, their gaseous discs
cannot be re-built. It is also unlikely that the remaining disc gas would
systematically gain angular momentum so as to regenerate a full exponential
disc. With these considerations, if $R_\text{str}(t_n) <
R_\text{str}(t_{n-1})$, the mass of cold gas removed by RPS at time $t_n$
can be shown to be  
\begin{equation}\label{eq:mrps}
  M_\text{RP}(t_n) = M_\text{cold}(t_n) \frac{f_s(t_n) -
    f_s(t_{n-1})}{1 - f_s(t_{n-1})},
\end{equation}
where $M_\text{cold}(t_n)$ is the mass of cold gas in the galactic
disc at the current snapshot but before RPS acts, and $f_s$ is given by
\begin{equation}
  f_s(t) = \left(1 + \frac{R_\text{str}(t)}{R_d} \right)
    \,\exp \left( -R_\text{str}(t)/R_d \right).
\end{equation}
This takes into account that if all the cold gas mass is between $R$~=~0
and $R$~=~$R_\text{str}$ in the form of a truncated exponential disc, then
the expression \eqref{eq:sigma0} for the central surface density of the
cold gas becomes 
\begin{equation}
  \Sigma_\text{0,cold}(t_n) = \frac{M_\text{cold}(t_n)}
        {2\pi \left[1-f_s(t_{n-1}) \right] R_d^2}\,.
\end{equation}

Conversely, if in a subsequent snapshot the galaxy experiences a lower RP,
$R_\text{str}(t_n)$ will be larger than the radius at which the remaining
gas is found, $R_\text{str}(t_{n-1})$. In such a case, we set
$R_\text{str}(t_n) = R_\text{str}(t_{n-1})$ and no gas is removed in that
time-step. 

In addition to RPS, gas could be removed from the galaxy by turbulent
stripping, caused by the generation of Kelvin-Helmholtz and/or
Rayleigh-Taylor instabilities at the interface between the cold gas
and the ICM. However, we do not consider this possibility as the
time-scale associated with the growth of such instabilities is
generally much longer than the RPS time-scale, as shown by
\citet{mccarthy2008} in their work \citep[see also][]{font2008}.

Finally, we would like to emphasize once again that RP is calculated for
every satellite galaxy in the entire simulation. We do not restrict the
calculation to galaxies residing within the most massive groups, and we do
not turn on the calculation at a selected redshift; RP will act on galaxies
whenever condition \eqref{eq:rps} is satisfied.


\section{Results}
\label{sec:results}
\subsection{Radial profiles and distributions of RP}
The procedure outlined in the previous section allows us to follow the
evolution of the properties of the intergalactic medium within DM~haloes,
obtaining for each snapshot values of the ambient density and relative
velocity for each satellite galaxy. In this section we discuss the
distributions and radial profiles of RP that we obtain for satellites in
the simulated clusters, and their evolution with time.

\begin{figure*}
  \centering
  \epsfig{file=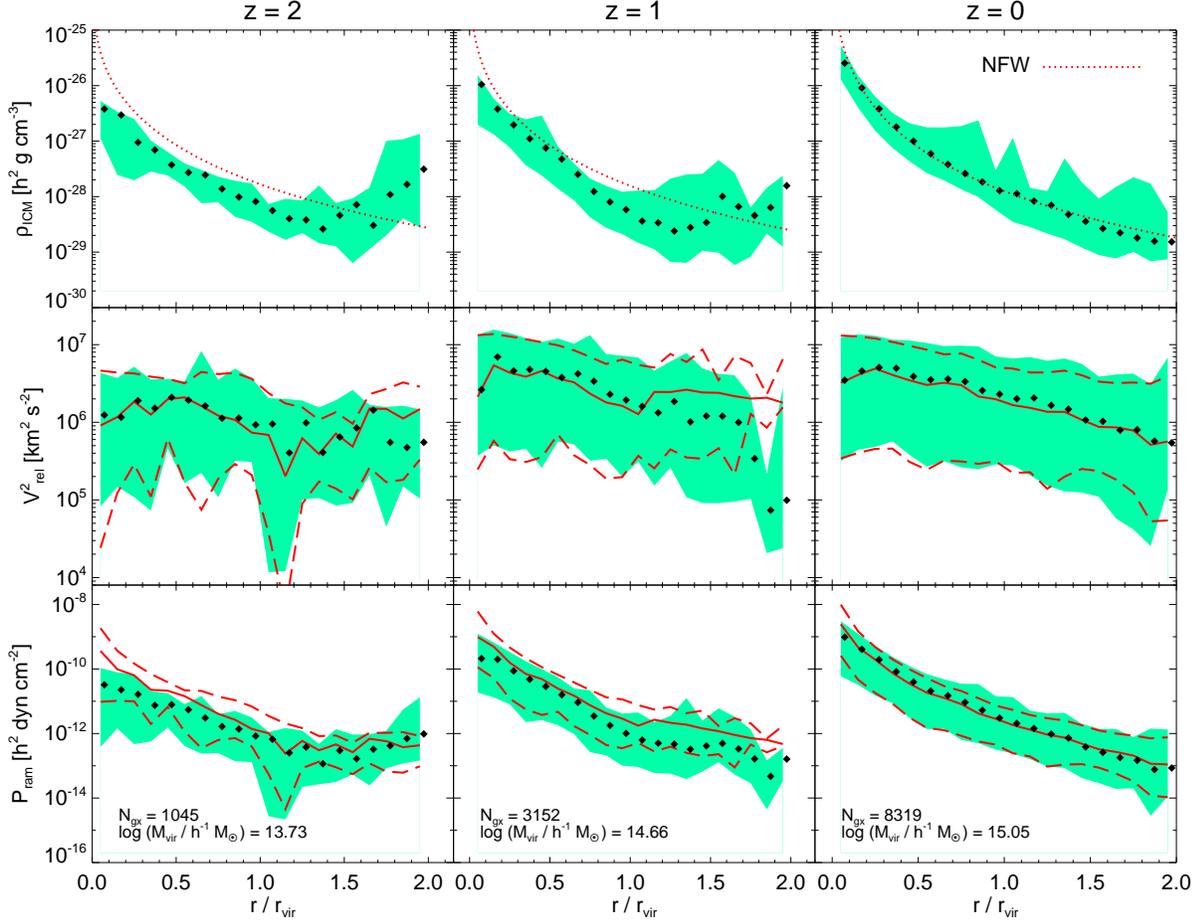, width=0.9\textwidth}
   \caption{Values of ICM density (top), galaxy velocity relative to
    the ICM (centre) and RP (bottom) as a function of the distance to
    the cluster centre, at the current positions of all satellite
    galaxies identified as members of the \fof~group corresponding to
    one of the G15 clusters. Results are shown for three different
    redshifts: $z$ = 2 (left), $z$~=~1 (centre) and $z$~=~0
    (right). Diamonds indicate median values in bins of 0.1~$r/$\rvir,
    and the shaded areas mark the regions enclosed by the 5th and 95th
    percentiles in each bin. The red dotted line in the top panels is
    a NFW density profile with the concentration $c$ of the halo at
    that redshift. The solid line in the central panels indicates the
    median velocity of the galaxies in the rest frame of the cluster,
    and the dashed lines enclose the 5th and 95th percentiles of that
    distribution. In the bottom panels, the solid line indicates the
    median RP obtained by multiplying the density given by the NFW
    profile by the rest-frame velocity; dashed lines denote the 5th
    and 95th percentiles.}
  \label{fig:icm}
\end{figure*}

\begin{figure*}
  \centering
  \epsfig{file=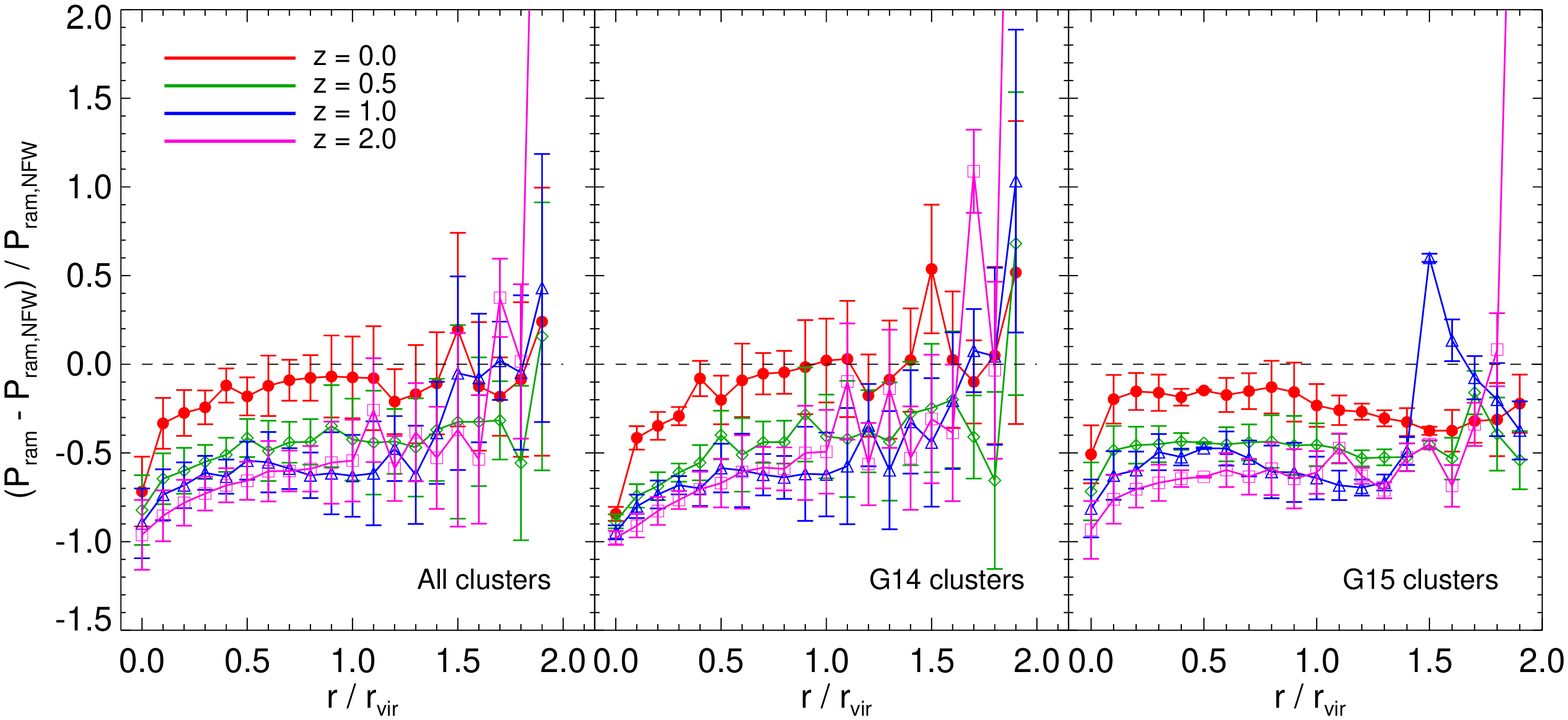, width=0.9\textwidth}
  \caption{Average difference between the RP determined from the gas
    particles in the simulation and the RP calculated as the product
    of a NFW density profile and the squared velocities of the
    satellite galaxies in the rest frame of the clusters. Results are
    shown for $z$~=~2, 1, 0.5 and 0, and for all the simulated
    clusters (left), the G14 clusters only (centre) and the G15
    clusters only (right).}
  \label{fig:rpdiff}
\end{figure*}

An interesting point is to evaluate how our RP model compares with previous
studies which assume analytic density profiles for the ICM
(\citealt{on2003,lanzoni2005}; BDL08). Hence, additionally, we
estimate the RP and the properties of galaxies assuming that the ICM gas is in
hydrostatic equilibrium following a NFW profile, re-scaled according to the
adopted baryonic fraction. These profiles are calculated using the
concentration parameter determined for the cluster at each
redshift. We will refer to this as the `analytic' method, and to our new
implementation described in Section \ref{sec:icmprop} as the
`gas-particles' method.

The dependence on clustercentric distance of the ICM density, relative
velocity (squared) of the satellites with respect to the ICM and the
resulting RP is plotted in Fig.~\ref{fig:icm} for three different
redshifts, where we show the results obtained for one of the G15 clusters
(taking into account the properties of the main cluster progenitor when
considering redshifts $z >$~0). In all other clusters we find a similar
behaviour, so we choose this one as representative of the general
trends. The median values obtained with the gas-particles method are
shown with black points, and green shaded areas mark the regions
enclosed by the 5th and 95th percentiles of the distributions. In
these plots we include all galaxies identified as satellites belonging
to the cluster \fof~halo at each selected redshift, and the ICM
properties are sampled at the positions of the galaxies. 

The top panels of Fig.~\ref{fig:icm} show the ICM density profiles
obtained. As the cluster grows in mass from
$\log($\mvir$/h^{-1}$~\msun$)\simeq$~13.7 up to 15.1 between $z$~=~2
and $z$~=~0, the median density in the cluster central region ($r
<$~0.5\rvir) increases about one order of magnitude. At the virial
radius, the increase between $z$~=~1 and $z$~=~0 is smaller. The
median density profile decreases monotonically at the present epoch
over four orders of magnitude, in the range 0~$< r / r_\text{vir}
<$~2. We compare these results with the ICM profiles determined for
the analytic method (shown with dotted lines in the top panels of
Fig.~\ref{fig:icm}). The match between the median density profile
determined from the simulation and the NFW analytic profile is
excellent at $z$~=~0; however, for $z \geq$~1, NFW profiles tend to
overestimate (underestimate) the density for $r <$~\rvir~($r
>$~1.5\rvir).

The median square velocities of the satellite galaxies relative to the
ICM are shown in the middle panels of Fig.~\ref{fig:icm}. The velocity
distribution of cluster galaxies is already established at $z
\simeq$~1 (at least for $r <$~1.5\rvir), as can be seen from the
similarities between the radial profiles at both $z$~=~1 and
$z$~=~0. There is a trend of increasing median relative velocity for
decreasing radius, with a large scatter at all radii. At $z$~=~2, the
velocity profile is flatter and the median velocities are lower than
for $z \leq$~1.

On any given cluster, the evolution of the RP experienced by satellite
galaxies for $z \leq$~1 depends mainly on the build-up of the ICM
density over time, as the cluster velocity profile is already
established at $z \simeq$~1. The bottom panels of Fig.~\ref{fig:icm}
show that RP increases approximately one order of magnitude at the
cluster centre between $z$~=~1 and $z$~=~0, consistent with the
increase in ICM density. Following the trend set by the density, the
RP stays at roughly the same level for 1~$< r /r_\text{vir} <$~1.5
between $z$~=~2 and $z$~=~1, and increases slightly between $z$~=~1
and $z$~=~0. For $r >$~1.5\rvir, the median RP seems to drop between
$z$~=~2 and $z$~=~1, and stays roughly similar for $z \leq$~1. This
happens because, in this region, the median density decreases over
time but the mean relative velocity of galaxies increases, thus
resulting in similar RP values. 

To determine the RP in the analytic method we also need an estimation of the
relative velocities of galaxies. If these are not determined using the gas
particles in the simulations, the alternatives are either to draw them at
random from an assumed distribution \citep{lanzoni2005} or to take the
velocity of the satellite (as given by the semi-analytic model) in the
rest frame of its central galaxy (BDL08). If the ICM is hydrostatic this
latter approach should be a good approximation, and this is what we use for
the analytic method.  The results of this are shown with red lines in the
middle panels of Fig.~\ref{fig:icm}. The agreement between the analytic
and gas-particles methods is again excellent at $z$~=~0, and also at
$z$~=~1 for $r \lesssim$~\rvir, for both the median (solid line) and the
scatter (dashed lines). At $z$~=~1, rest-frame velocities tend to be larger
for $r >$~\rvir~than the velocities relative to the gas particles.

The analytical distribution of RP that results of multiplying the median
density given by the NFW profile by the square of the rest-frame velocities
is shown with lines in the bottom panels of Fig.~\ref{fig:icm}. As
expected from the very good agreement between both approaches, both
for the density and the relative velocity at $z$~=~0, the RP profiles are very
similar at the present epoch. The agreement is not so good for $z \geq$~1
where the shape of the RP profiles are similar but in the analytical model
the mean values and the levels of the percentiles shift to higher
values. According to the results of our self-consistent numerical approach,
the analytic method appears to overestimate the RP over most of the range
considered.

To quantify this latter point, in Fig.~\ref{fig:rpdiff} we plot the
average difference between the RP determined from the gas-particles and the
analytic methods, normalized to the analytical value of RP. Results are
shown in different colours for $z$~=~2, 1, 0.5 and 0. The left-hand
panel shows the result of averaging over all the simulated clusters,
and error bars show the 1$\sigma$ cluster-to-cluster scatter. From
this plot we can clearly see that the analytic method always
overestimates the value of RP for $r \lesssim$~1.5\rvir~(although at
$z$~=~0 and for $r \gtrsim$~0.4\rvir~it could be said that both models
agree within the errors). The difference between the two models
increases with redshift; this trend is clear within \rvir, but becomes
noisier outside of it (perhaps due to the smaller number of cluster
galaxies in these regions). These trends persist if we consider
separately the G14 and G15~clusters (middle and right-hand panels in
Fig.~\ref{fig:rpdiff}, respectively).

\begin{figure}
  \centering
  \epsfig{file=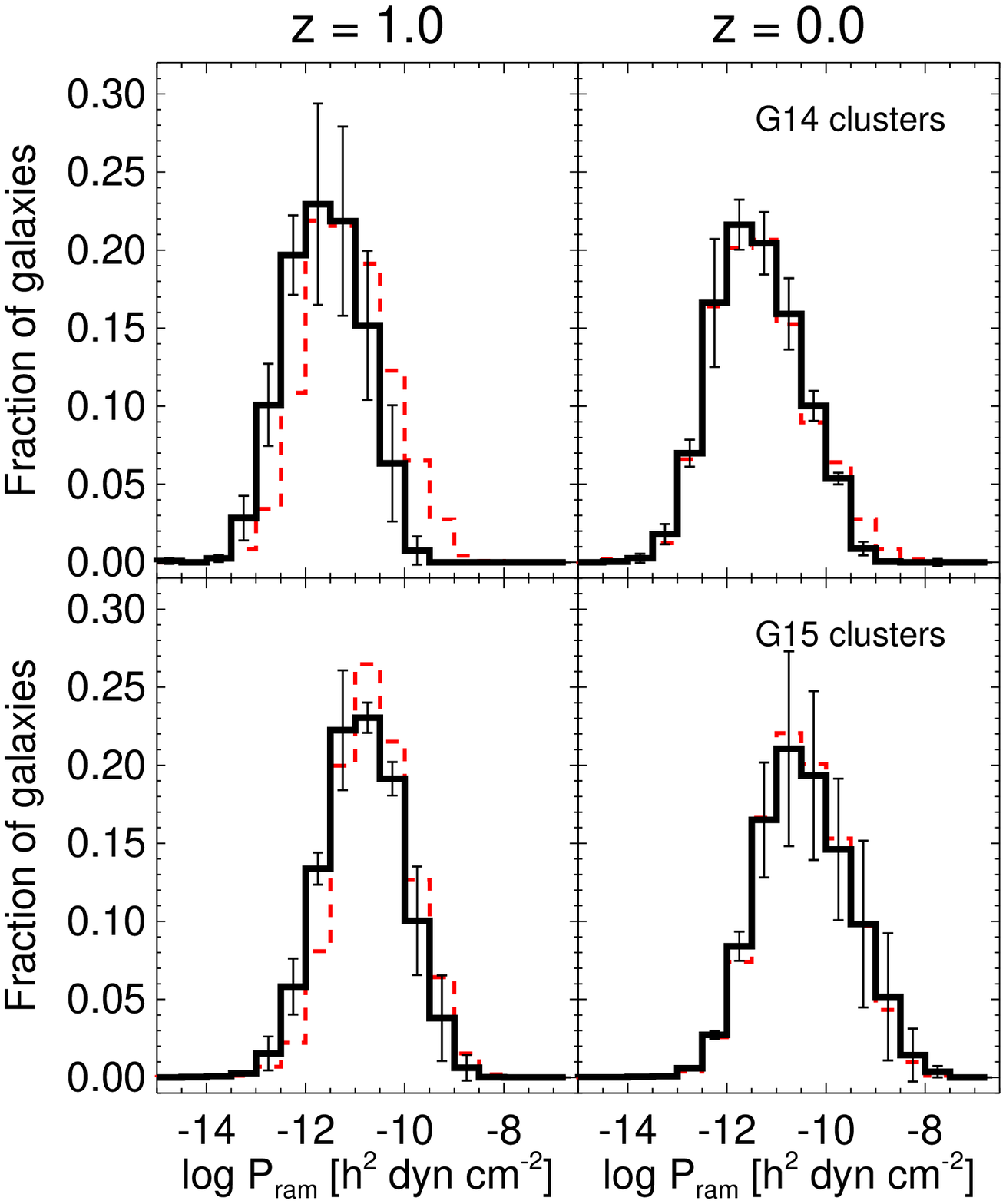, width=0.45\textwidth}
  \caption{Histograms of the RP experienced by the galaxies within
    \rvir~in the simulated clusters at $z$~=~1 (left) and $z$~=~0
    (right), obtained from the gas-particles method (black lines) and
    the analytic method (dashed lines). Results for the G14 and
    G15~clusters are shown in the top and bottom panels,
    respectively. Error bars denote the~1$\sigma$~cluster-to-cluster
    scatter and for clarity are shown only for the gas-particles method.}
\label{fig:pramdist}
\end{figure}

The distribution of RP values experienced by satellite galaxies within
\rvir~in our simulated clusters is shown in Fig.~\ref{fig:pramdist}
for the G14~clusters (top) and G15~clusters (bottom) at redshifts $z$~=~1
(left) and $z$~=~0 (right). We compare results obtained from the
gas-particles method (histograms in black lines) and the analytic method
(histograms with dashed lines). All the distributions can be well
fitted by Gaussian functions, and the result of the fits we obtained for
these histograms is shown in Table~\ref{tab:pramfits}. The mean value of
RP obtained from both the gas-particles and analytic methods are higher for
the more massive clusters, a feature that is also present at higher
redshifts. We mentioned before that for a given cluster the increase of RP
for $z \lesssim$~1 is due to the increase in ICM~density, but the
difference between the RP values for different cluster masses is due to a
combination of increased density and relative velocity. On average, the
ICM~density within \rvir~in G14~clusters can be as high as 75 per cent
of the density at same radii in G15~clusters. The rest of the
difference is due to the lower relative velocity of satellites in the
smaller clusters (within \rvir, the mean velocity in G14~clusters is
$\sim$25 per cent of the mean $v$ in G15~clusters). 

The range of RP values we obtain at $z$~=~0 for both models is similar
to that found by BDL08, who calculate values of RP for clusters of
masses comparable to ours, extracted from the \citet{delucia2007}
semi-analytic catalogue. BDL08 assume an ICM described either by an
isothermal or a \citet[][hereafter KS01]{ks2001} model, in
hydrostatical equilibrium within a NFW halo. For the case of the
isothermal model, BDL08's distribution of RP is fairly skewed, with a
sharp cut-off at higher pressures; for the KS01 model they obtain a
much more symmetrical distribution (see their figs 1 and 2,
respectively). Comparing the shapes of our local RP distributions,
given in the right-hand panels of Fig.~\ref{fig:pramdist}, with those
of BDL08, we find better agreement with their results for the KS01
model. The local mean RP values obtained from our gas-particles method
(see Table~\ref{tab:pramfits}) are also in good agreement with those
determined by BDL08 with the KS01 model, who find
10$^{-11.3}$~dyn~cm$^{-2}$ for their set of 10$^{14}$~\msun~clusters,
and 10$^{-10.7}$~dyn~cm$^{-2}$ for their 10$^{15}$~\msun~clusters.

\begin{table*}
  \caption{Best-fitting parameters of Gaussian functions to the RP
    histograms of Fig.~\ref{fig:pramdist}. $\langle \log P_\text{ram}
    \rangle$ and $\sigma_\text{ram}$ denote the mean and dispersion of each
  distribution, respectively.}
  \label{tab:pramfits}
  \begin{center}
    \leavevmode
    \begin{tabular}{c c c c c c} \hline \hline
      & \multicolumn{2}{c}{G14 clusters} & \multicolumn{2}{c}{G15 clusters}
       & Model \\
       $z$ & $\langle \log P_\text{ram} \rangle$ & $\sigma_\text{ram}$ & 
                 $\langle \log P_\text{ram} \rangle$ & $\sigma_\text{ram}$ \\ 
       & \multicolumn{2}{c}{[$h^{2}$ dyn cm$^{-2}$]} & 
        \multicolumn{2}{c}{[$h^{2}$ dyn cm$^{-2}$]} \\ 
      \hline
      1.0 & -11.64$\pm$0.08 & 0.73$\pm$0.05 & -10.91$\pm$0.03 &
       0.76$\pm$0.02 & Gas-particles\\
          & -11.14$\pm$0.07 & 0.84$\pm$0.05 & -10.63$\pm$0.05 &
       0.75$\pm$0.03 & Analytic\\
      0.0 & -11.33$\pm$0.03 & 0.87$\pm$0.02 & -10.51$\pm$0.09 &
       0.83$\pm$0.04 & Gas-particles\\
          & -11.18$\pm$0.05 & 0.96$\pm$0.03 & -10.47$\pm$0.07 &
       0.81$\pm$0.03 & Analytic\\
      \hline
    \end{tabular}
    \end{center}
 \end{table*}

\begin{figure}
  \centering
  \epsfig{file=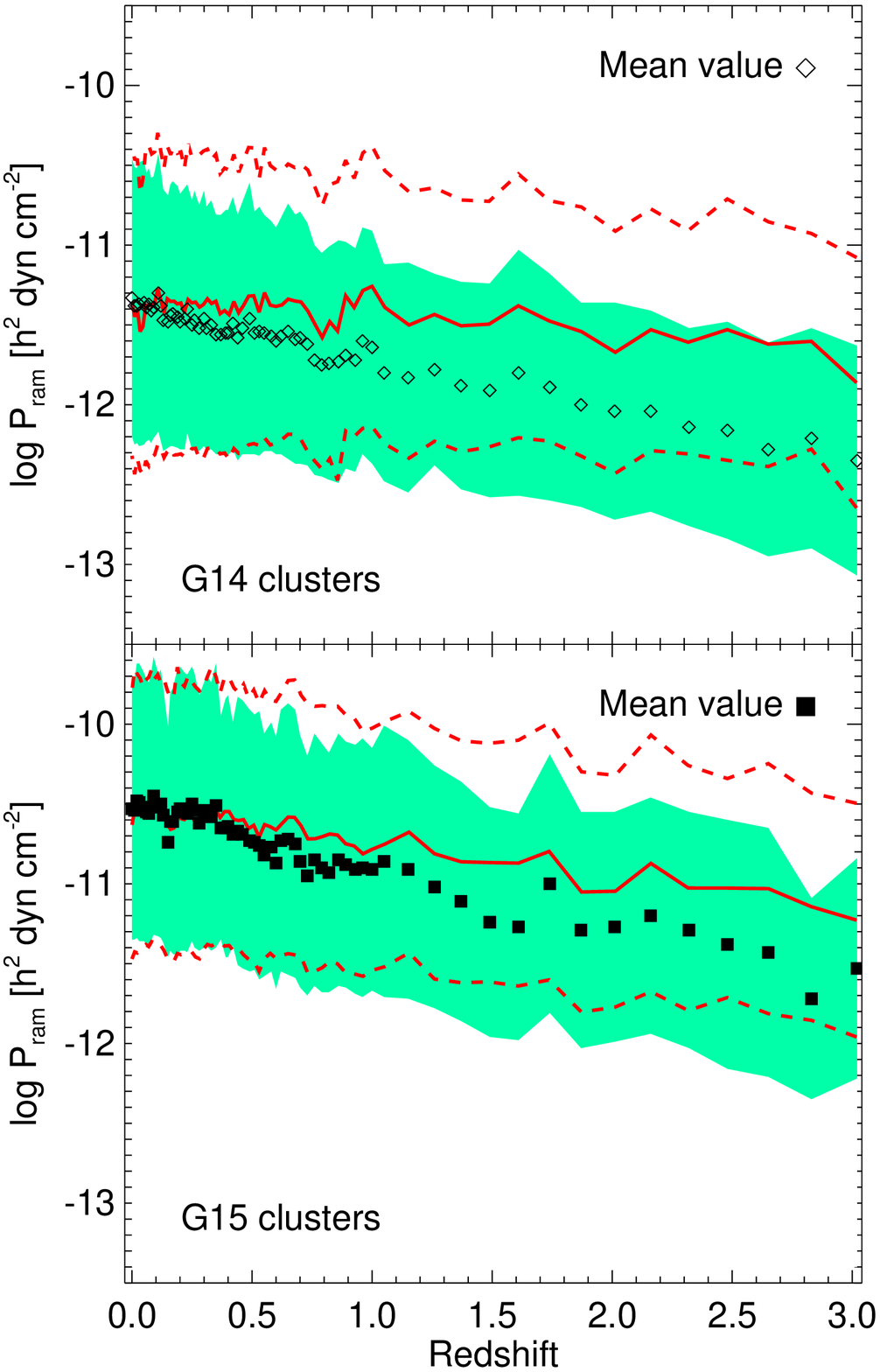, width=0.45\textwidth}
  \caption{Mean ram pressure exerted on the galaxies within \rvir~of
    the simulated clusters as a function of redshift. Top and bottom panels
    show results for the G14 and G15 clusters, respectively. Open diamonds
    and filled squares indicate the mean value of RP for the gas-particles
    method and the shaded areas denote the regions within 1$\sigma$ of the
    mean. Solid red lines indicate the mean RP for the analytic method,
    and the dashed red lines enclose the region within 1$\sigma$ of the
    mean.}
  \label{fig:rpvsz}
\end{figure}

As already noticed from Fig.~\ref{fig:icm}, we can see again in
Fig.~\ref{fig:pramdist} that the distributions obtained with both
methods agree very well at the present epoch, but not so for higher
redshifts. For both sets of clusters, the mean value of RP obtained
from the gas-particles method for galaxies within the virial radius
increases approximately half an order of magnitude from $z$~=~1 to the
present. However, in the analytical calculation, the mean values at
both redshifts are much more similar (in the case of the G14~clusters,
if errors are taken into account then the mean values can be
considered equal). This fact, combined with the results shown in
Fig.~\ref{fig:rpdiff}, indicates that the RP in the clusters does not
grow over time in the same way in both models. 

To better visualize this, we plot in Fig.~\ref{fig:rpvsz} the mean RP
for galaxies within \rvir~as a function of redshift, for both sets of
clusters. Mean RP values given by the gas-particles method are shown with
black symbols, and the shaded areas indicate the regions within 1$\sigma$
of the mean. Solid lines indicate the mean values obtained from the
analytical calculation, with dashed lines enclosing the regions within
1$\sigma$ of the mean. For both models, the difference between the
mean RP values of both cluster sets at a given $z$ is always about one
order of magnitude, being larger in the more massive clusters, as expected. 

In the simulations by \citet{roediger2006}, values of RP of order
10$^{-12}$~dyn~cm$^{-2}$ were called weak, 10$^{-11}$~dyn~cm$^{-2}$ medium
and 10$^{-10}$~dyn~cm$^{-2}$ strong. \citet{roediger2006} find that a
spiral galaxy with mass $\sim$~2~$\times$~10$^{11}$\msun~subject to strong
RP will typically lose all of its gas within~$\sim$~50~Myr, medium RP will
remove approximately half of the gas within~$\sim$~200~Myr, and weak
RP removes relatively small amount of gas (the actual gas loss will
depend on the structure of the gaseous, stellar and DM
components). Fig.~\ref{fig:rpvsz} shows that for the gas-particles
model the majority of satellite galaxies in the massive G15~clusters are
experiencing medium level RP already at $z \sim$~2, and at $z
\lesssim$~0.5 about 20 per cent of satellites experience strong levels
of RP at any given time. In the case of the G14~clusters, the mean RP
is between medium and weak for $z \lesssim$~1.5, and few galaxies ever
experience strong levels of RP; only at $z \sim$~1 does a significant
fraction of satellites begin to experience medium levels of RP.

In the analytic model, at higher redshifts ($z \gtrsim$~1), the mean RP
for the G15~clusters reaches medium values and a significant fraction of
satellite galaxies already experience medium-to-strong levels of RP at $z
\sim$~2. A similar situation occurs for the G14~clusters, with the values
of RP shifted to the range of weak-to-medium levels, the difference between
the two models being larger for higher redshifts. In both models the
scatter around the mean is large, spanning one order of magnitude at each
side of the mean. The dispersion in the distributions also grows slightly
with time, as can be better appreciated from Table~\ref{tab:pramfits},
where we can see an increase in $\sigma$ of about 10 per cent between
$z$~=~1 and 0, for both sets of clusters. This may arise as a result
of the larger number of galaxies in clusters at the present time
(about three times more galaxies than at $z$~=~1).

Although the results for both models agree at $z \sim$~0 for the different
sets of clusters, the different evolution of RP over time is clear. While
the mean RP values obtained from gas-particles method increase almost
exponentially with redshift, the evolution of the mean RP in the analytic
method is flatter, especially so in the case of the G14~clusters. 
The lack of strong evolution of the mean RP values for G14~clusters is a
consequence of the overestimation of RP by the analytic model for the inner
parts of the clusters ($r<0.5$\rvir), even at $z \sim$~0 (see middle panel
of Fig.~\ref{fig:rpdiff}). This is because these clusters, and their
progenitors at higher redshift, are galaxy group-sized systems of virial
masses in the range $\sim$~10$^{13}-10^{14}\,h^{-1}$~\msun, whose ICM
distributions are not so well described by a NFW profile. There is also a
clear but smaller difference between the models for the case of the
G15~clusters. For $z \lesssim$~0.5, both models agree fairly well; this can
be due to the fact that these clusters are already massive systems at $z
\sim$~2, and so their ICM is better described by a NFW profile (see radial
profiles for different cluster masses in the top panels of
Fig.~\ref{fig:icm}).

\begin{figure*}
  \centering
  \subfigure {
    \epsfig{file=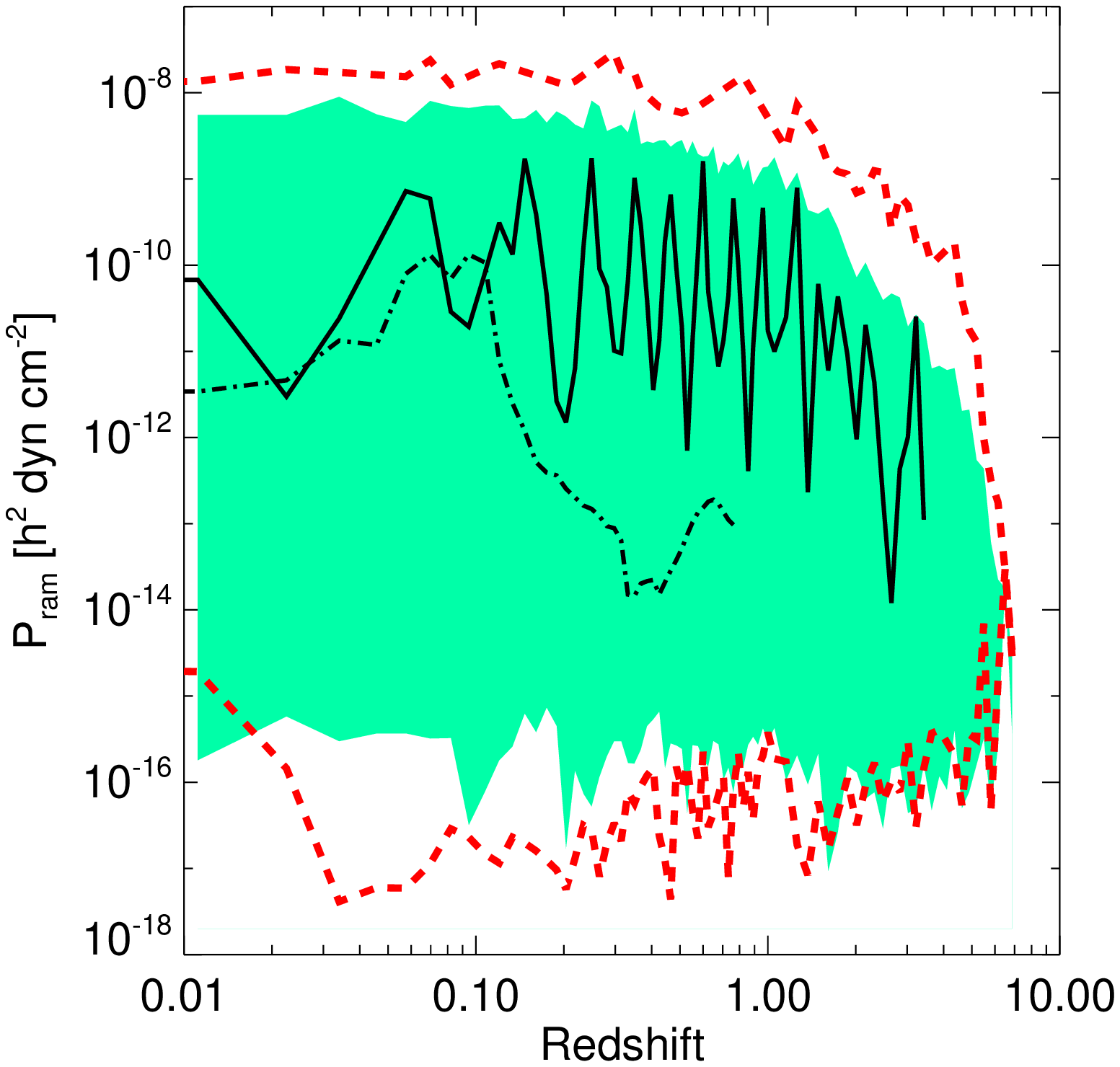, width=0.32\textwidth}
  }
  \subfigure {
    \epsfig{file=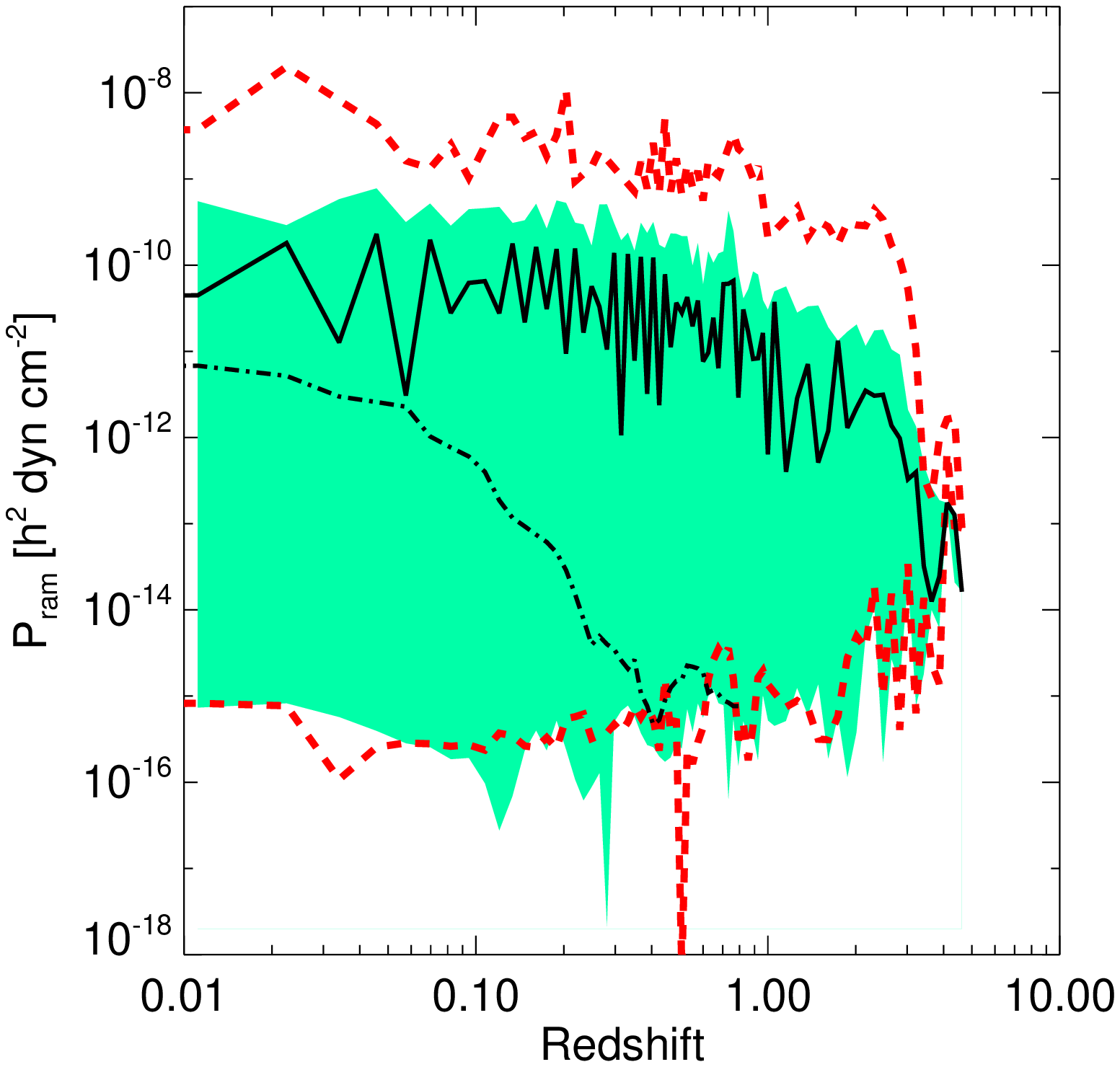, width=0.32\textwidth}
  }
  \subfigure {
    \epsfig{file=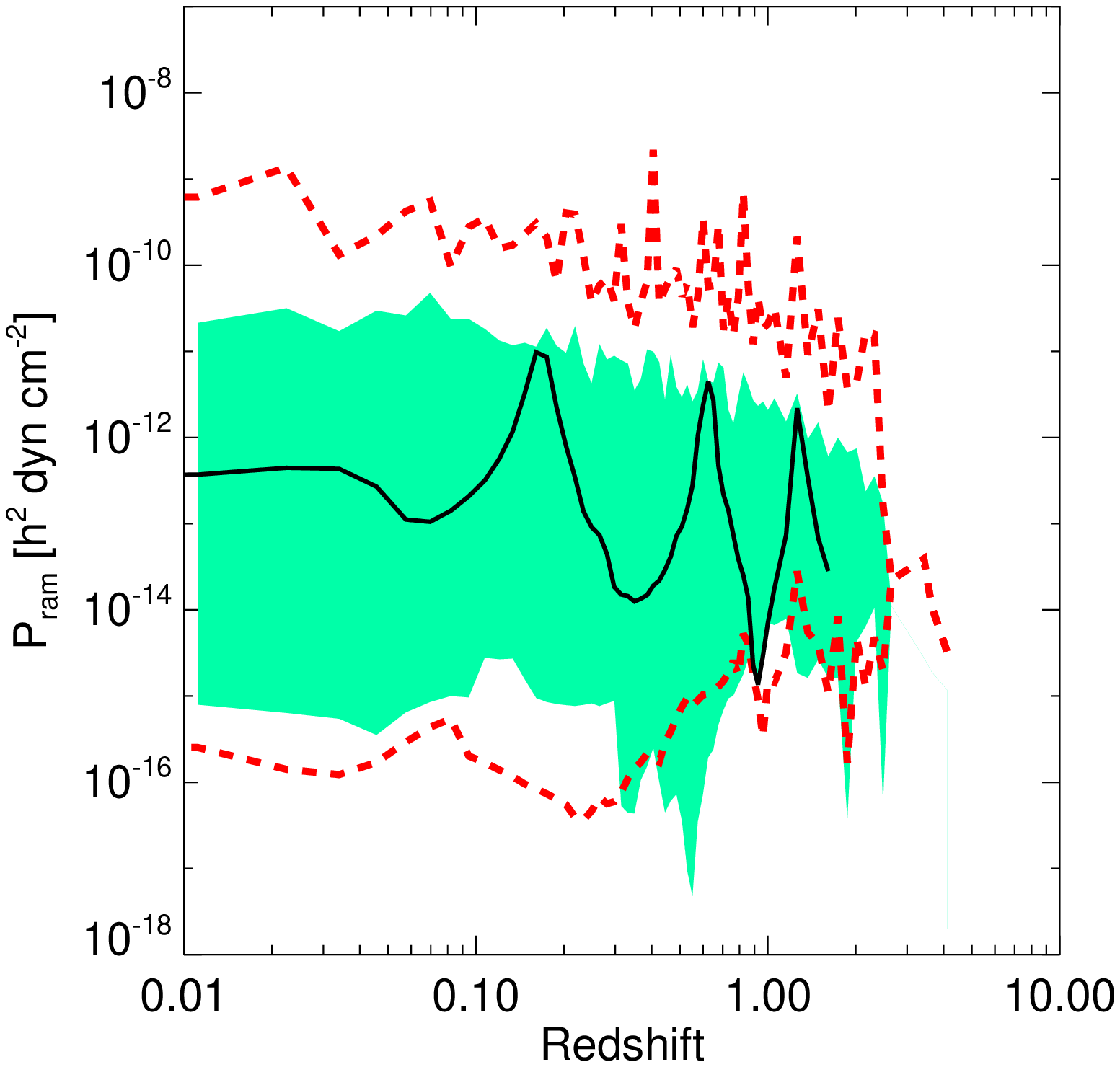, width=0.32\textwidth}
  }
  \caption{RP histories of satellite galaxies in one of the G15
    clusters (left), one of the G14 clusters (centre) and in one group
    with \mvir~$\simeq$~10$^{13}\,h^{-1}$\msun~(right) which is
    contained in the simulation box of one of the G15 clusters. Solid
    black lines correspond to a selected type~2 satellite, and black
    dot-dashed lines to a type~1 satellite. Shaded areas mark the
    regions enclosed by the maximum and minimum RP reached by all the
    galaxies present in the cluster at $z$~=~0 as determined from the
    gas-particles method; dashed thick lines denote the maximum and
    minimum values of the RP calculated with the analytic method.}
  \label{fig:rphist}
\end{figure*}

In Fig.~\ref{fig:rphist} we plot RP histories given by our gas-particles
method for randomly selected satellite galaxies of type~1 (black
dot-dashed lines) and type~2 (black solid lines). Results are shown
for one of the G15 clusters (left) and one of the G14 clusters
(centre). For comparison, the right-hand panel of
Fig.~\ref{fig:rphist} shows the RP evolution for galaxies within one
halo with virial mass $\simeq$~10$^{13}\,h^{-1}$~\msun~extracted from
the simulation box of one of the G15 clusters. There are differences
between the RP histories of different types of satellites. For type~2
satellites, the RP typically oscillates, increasing on average;
whereas for type~1 galaxies, the typical history is one of
monotonically increasing RP. This difference in the RP histories is
due to the way in which positions and velocities of galaxies are
assigned in the code, being more reliable for type~1 galaxies. As
mentioned in Section~\ref{sec:sag} the position and velocity of a
galaxy are assigned by tracking the most-bound DM particle of its host
DM subhalo. In the case of type~1 galaxies this subhalo still exists
and is affected by dynamical friction, but for type~2 galaxies it has
been destroyed by tidal forces and we are then tracking a single DM
particle. The regions enclosed by the maximum and minimum RP values
reached by all the galaxies that are present at $z$~=~0 in the
selected halo are depicted by dashed lines and shaded areas for the
analytic and gas-particles methods, respectively. Once again, we see
differences between the models; satellites in the analytical model
reach higher values of RP in all three cases, and the difference
between both methods increases with decreasing halo mass. 

\subsection{Effects of ram pressure stripping}
\label{sec:rpseff}
After analysing the evolution of the distributions of RP values and the
dependence on the cluster mass obtained from two different models, we now
focus on the values given by the gas-particles method and explore the
influence of RPS on galaxy properties. The new estimation of the size
of galactic discs, described in Section~\ref{sec:sizes}, affects the
results of the disc instability process implemented in \sag. This
calls for a retuning of the free parameters of \sag~in order to retain
a good fit to observed properties such as the $b_J$- and $K$-band
cluster luminosity functions and the BH-bulge mass relation, among
others. Three parameters change with respect to those used by LCP08
(see the descriptions of the model parameters in LCP08). Specifically,
the parameters $\kappa_\text{AGN}$ and $f_{\rm BH}$, which control the
rate of gas accretion onto the central BH, increase from
2.5~$\times$~10$^{-4}$~\msun~yr$^{-1}$ to 10$^{-3}$~\msun~yr$^{-1}$,
and from~0.015 to 0.04, respectively; the disc instability threshold
$\epsilon_\text{disc}$ decreases from~1.1 to 0.85. This recalibration
is done without considering RPS. We then run three sets of models: one
without and two with RPS, using the same set of parameters in all
cases in order to evaluate the effect of RPS on galaxy properties. In
the following, we will use \sag~to refer to the semi-analytic model
without RPS, and \sagrp~and \sagrpa~will denote the models where RP
values are determined using the gas-particles or the analytical
methods, respectively.

\begin{figure}
  \centering
  \epsfig{file=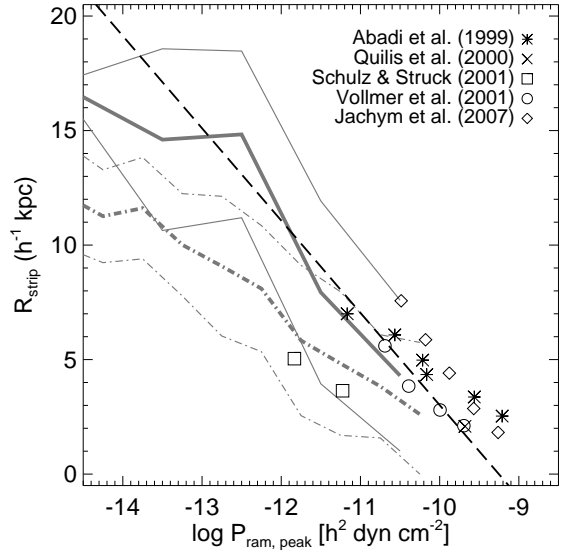, width=0.45\textwidth}
  \caption{Stripping radius $R_\text{str}$ as a function of
      the peak~RP experienced by disc satellite galaxies
      ($M_\text{bulge}/M_\text{stellar} \leq$~0.95). Thick dot-dashed
      and solid lines denote the mean $R_\text{str}$ for satellites
      with 10$^{10}\,h^{-1} M_\odot \leq M_\text{stellar}
      <$~7~$\times$~10$^{10}\,h^{-1} M_\odot$ and with
      7~$\times$~10$^{10}\,h^{-1} M_\odot \leq M_\text{stellar}
      <$~2~$\times$~10$^{11}\,h^{-1} M_\odot$, respectively, and
      thinner lines enclose the regions within 1$\sigma$ of the
      mean values. Points show the results of different 
      N-body/hydrodynamical simulations of RPS in single galaxies. The
      dashed line shows the analytical estimate of
      equation~\eqref{eq:rstrip} for a Milky Way-class galaxy, with
      surface densities of stars and gas taken from \citet{flynn2006}.}
  \label{fig:rs}
\end{figure}

As a test of the performance of the model, Fig.~\ref{fig:rs} shows the
stripping radius of disc satellite galaxies for the \sagrp~model as a
function of the peak RP that they experience. Results for the mean
stripping radius and 1$\sigma$ scatter are plotted for disc satellite
galaxies in two different mass ranges: 10$^{10}\,h^{-1} M_\odot \leq
M_\text{stellar} <$~7~$\times$~10$^{10}\,h^{-1} M_\odot$ (dot-dashed
lines) and 7~$\times$~10$^{10}\,h^{-1} M_\odot \leq M_\text{stellar}
<$~2~$\times$~10$^{11}\,h^{-1} M_\odot$ (solid lines). We compare this
with the results of several N-body/hydrodynamical simulations
specially focused on the study of RPS in individual galaxies (denoted
by different symbols). These simulations employ different procedures
to compute the hydrodynamics (SPH:~\citealt{abadi99},
\citealt{ss2001}, \citealt{jachym2007};
Eulerian:~\citealt{quilis2000}, \citealt{roediger2005}; sticky
particles:~\citealt{vollmer2001}). All the simulations consider Milky
Way-class galaxies entering a dense environment, except for
\citet{ss2001} who study a smaller galaxy with $M_\text{stellar}
\simeq$~6~$\times$~10$^{10}\,h^{-1}$~\msun. \citet{jachym2007} run
simulations for different types of clusters; here we select the
results for their standard cluster model. The dashed line shows the
analytical estimate for $R_\text{str}$ of Eq.~\eqref{eq:rstrip} for a
Milky Way-like galaxy, with surface densities for stars and gas taken
from \citet{flynn2006}. There is a good agreement between the GG72
estimate and the different simulations (for clarity we do not plot the
results from \citealt{roediger2005}, as they are very similar to the
GG72 line).

Even though a comparison between our method and the detailed
simulations is not direct, since the initial conditions are not
necessarily the same, the general trend seen in the detailed
simulations is reproduced in our model for galaxies of similar stellar
masses. We note that the comparison does not include those satellite
galaxies from our simulations whose cold gas is completely stripped
($R_\text{str}$~=~0). There are no such galaxies for $\log
P_\text{ram,peak} \lesssim$~-11.5, and the majority of fully stripped
galaxies~($\gtrsim$80 per cent) are found at $\log P_\text{ram,peak}
\gtrsim$~-10.5.

\begin{figure*}
  \centering
  \epsfig{file=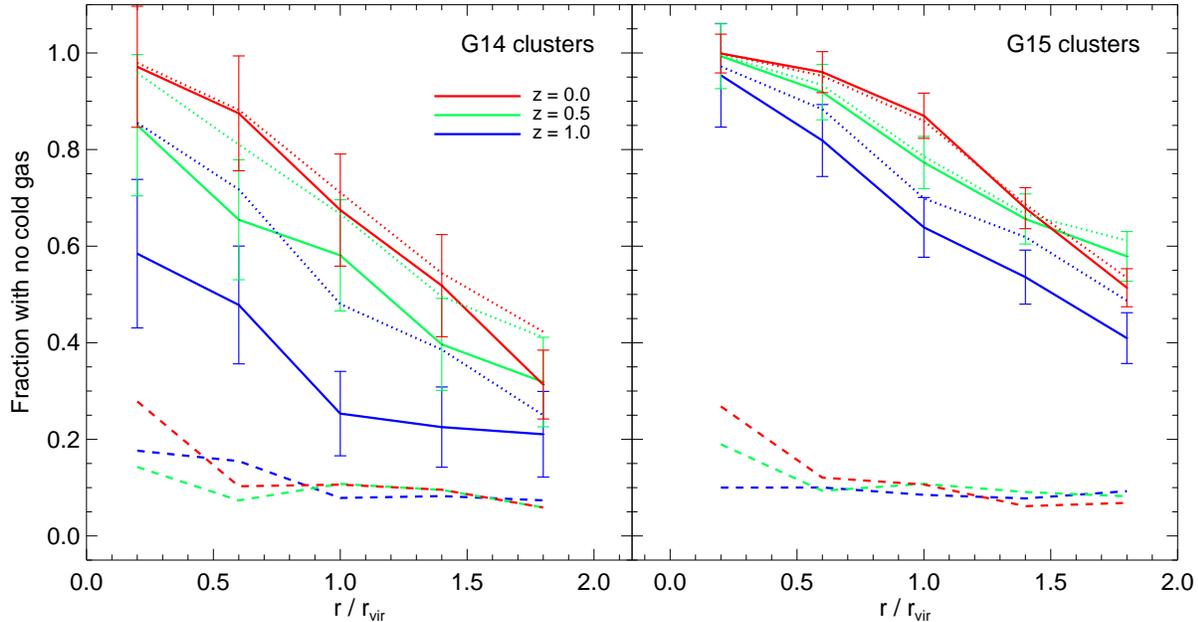, width=0.9\textwidth}
  
  \caption{Fractions of galaxies that have lost all of their cold gas
    as a function of distance to the cluster centre, for G14 clusters
    (left) and G15 clusters (right). Results are shown for three
    different redshifts: $z$~=~1, 0.5 and 0, and for three models:
    \sag~(dashed lines), \sagrp~(solid lines), which takes into
    account RP values estimated from the gas-particles method, and
    \sagrpa~(dotted lines), where the RP is calculated with the
    analytic method. All galaxies within 2\rvir~of the simulated
    clusters and with stellar mass greater than 10$^9\,
    h^{-1}$~\msun~are included in this plot. Galaxies are binned in
    0.4$r/$\rvir, and error bars are only shown on the \sagrp~model
    for clarity.} 
  \label{fig:fracnogas}
\end{figure*}

One of the galaxy properties most directly affected by RPS is the cold
gas content. Fig.~\ref{fig:fracnogas} shows the fraction of galaxies
that have lost all of their cold gas ($f_\text{no-gas}$) as a function
of clustercentric distance, for G14 and G15~clusters. To construct
this plot, we select galaxies within 2\rvir~of the simulated clusters
and with stellar mass~$\geq$~10$^9\, h^{-1}$~\msun. For this plot all
galaxies within \rvir~are considered, not only members of the main
\fof~group; so we are including members of outlying groups which are
infalling into the main cluster. The fractions of gas-depleted
galaxies are determined for the three different semi-analytic models
considered, which are represented by different line types. Different
colours identify the redshifts at which results are shown. For the
\sag~model, $f_\text{no-gas}$ is very low; only in the innermost
regions ($r \lesssim$~0.2\rvir) the gas fractions increase from
$f_\text{no-gas} \approx$~0.2 to 0.3 in the redshift interval 1~$< z
<$~0, regardless of cluster mass. These galaxies are those that have
spent enough time within the cluster to gradually consume their cold
gas reservoir; recall that cooling flows are suppresed in satellite
galaxies due to strangulation. In \sag,~$f_\text{no-gas}$ has an
almost flat radial distribution beyond 0.5\rvir~(dashed lines in
Fig.~\ref{fig:fracnogas}). 

When RPS is acting, almost all galaxies in the innermost regions of
clusters lose their cold gas by $z$~=~0, for both sets of clusters. The
fractions of gas-depleted galaxies in the \sagrp~model decrease for
larger distances to the cluster centre, but they remain significant at
$r\simeq$~2\rvir, being larger for G15~clusters
($f_\text{no-gas}\simeq$~0.5) than for G14~ones
($f_\text{no-gas}\simeq$~0.3). The lower values of $f_\text{no-gas}$
in the outskirts of G14~clusters in the models that include RPS are
also consistent with the weak-to-medium values that RP takes in these
clusters. This monotonic decrease of $f_\text{no-gas}$ with
clustercentric distance is a direct consequence of the behaviour of
the radial profile followed by RP values (see bottom panels of
Fig.~\ref{fig:icm}).

The evolution of the cold gas content of galaxies in the \sag~model is
very similar for clusters of different masses; however, this is not
the case for galaxies in models which include RPS. The difference of
over half an order of magnitude between the mean RP values for G14 and
G15~clusters at different redshifts, which is clearly visible in
Fig.~\ref{fig:rpvsz}, drives a rather different evolution of the
fraction of galaxies devoid of cold gas. In the \sagrp~model, for
G15~clusters the situation at $z$~=~0 regarding the cold gas content
of galaxies has been practically established at $z$~=~0.5, and it does
not evolve much since $z$~=~1. Although at this latter redshift the
fractions are $\approx$~10-20 per cent lower than at $z$~=~0 at all
radii, they are quite high ($f_\text{no-gas}\gtrsim$~0.5) even in the
cluster outskirts. Conversely, for the case of the G14~clusters we see
a stronger evolution in the fractions of gas-depleted galaxies in the
same redshift interval. This reflects a more gradual removal of the
cold gas of satellites as a result of the lower values of RP in these
clusters. In the outskirts of the G14~clusters $f_\text{no-gas}
\sim$~0.3, lower than for G15~clusters but still much larger than in
the \sag~model.

The results described above clearly show a dependence of the evolution
of the fraction of gas-depleted galaxies on cluster mass. This trend
is different for the analytic estimation of RP used in the \sagrpa~model,
shown by dotted lines in Fig.~\ref{fig:fracnogas}. Both for the G14
and G15~clusters, the final values of $f_\text{no-gas}$ are reached at
higher redshift than for the \sagrp~model. For instance, in the case
of the G14~clusters, the value of $f_\text{no-gas}$ in the
\sagrpa~model at $z$~=~1 is already as high as the value reached for
the \sagrp~model at $z$~=~0.5. This behaviour is the result of the
much higher values of RP at $z >$~0 in the \sagrpa~model (see
Fig.~\ref{fig:rpvsz}).

In the models that include RPS, the higher fractions of gas-depleted
galaxies found at $z$~=~0 in the cluster outskirts, for both sets of
clusters, indicate that during cluster assembly RPS has an important
effect on the galaxies contained in the smaller subgroups that are
being accreted by the cluster; this is what has been called `pre-processing'
\citep[e.g.][]{fujita2004,mihos2004}. This idea is also inferred from
observational evidence of compact group of galaxies falling into the
massive galaxy clusters Abell~1689 ($z \sim$~0.18), Abell~2667 ($z
\sim$~0.23) and Abell~1367 ($z \sim$~0.02)
\citep{cortese2006,cortese2007}; galaxies within these groups show
extended tails of ionized gas, that can be interpreted as the
signature of RPS. We then find that RPS is a process that
significantly affects the cold gas content of galaxies and the rate at
which it is removed, which depends on cluster mass. Consequently, the
RPS experienced by satellite galaxies might enhance differences in the
star formation history and the colour evolution on haloes of different
masses; this will be explored in detail in a forthcoming paper.


\section[]{CONCLUSIONS}
\label{sec:conclu}
We develop a new method to describe the effects of RPS on galaxy
properties which works within the \sag~semi-analytic model of galaxy
formation and evolution, in combination with cosmological
hydrodynamical $N$-body/SPH simulations of galaxy clusters. RPS is
implemented in \sag~adopting the GG72 criterion. The novel feature of
our implementation is that the kinematical and thermodynamical
properties of the hot gas responsible for RP are obtained from the gas
particles of the SPH simulations (gas-particles method). This results
in a more self-consistent estimation of the RP experienced by
satellite galaxies.

We compare our results with those obtained from an analytic estimation
of RP, which considers a NFW density profile for the hot gas contained
within DM~haloes, re-scaled according to the adopted baryonic fraction
(analytic method). We analyse the dependence on clustercentric
distance and redshift of the RP values given by both methods, and
evaluate the influence of the environment on the behaviour of these
distributions. The RPS method discussed in this work can be adapted to
work with any kind of cosmological simulation that includes gas
physics, regardless of the particular numerical scheme used for the
hydrodynamical calculations. However, the specific results from its
application might depend on the details of the numerical
implementation. For example, the SPH method utilises an artificial
viscosity term to properly capture hydrodynamic shocks, and this
viscosity can artificially suppress turbulence in the ICM
\citep{dolag05,agertz2007}. The simulations we have used include the
standard SPH artificial viscosity. The use of a lower-viscosity
formulation of SPH, or a grid-based method, may result in an increased
level of ICM turbulence. This could result in the distribution of the
velocities of galaxies relative to the gas deviating further from the
velocities calculated by assuming a hydrostatic ICM.

We have selected from the simulations considered two sets of
cluster-sized haloes, with masses \mvir~$\simeq$~10$^{14} \, h^{-1}
M_\odot$ (G14 clusters) and \mvir~$\simeq$~10$^{15} \, h^{-1} M_\odot$
(G15 clusters). The RP~values obtained by the gas-particles and
analytic methods are used by the RPS process implemented in the
semi-analytic model, resulting in the models \sagrp~and \sagrpa,
respectively. These models provide a galaxy population affected by
RPS; their results are compared with those obtained from the standard
\sag, which does not take RPS into account, in order to evaluate the
effect of RPS on the cold gas content. We summarize our main results. 

\begin{enumerate}
\item The RP estimated from the gas-particles method increases
  approximately one order of magnitude at the cluster centre between
  $z$~=~1 and 0, consistent with the increase in ICM density, since
  the cluster velocity profile is already established at $z$~=~1. Median RP
  values do not evolve much in the outskirts of the cluster
  ($r/r_\text{vir} >$~1), which is a consequence of the combined behaviour
  of the median ICM density and the relative velocities, whose radial
  distributions decrease and increase with time, respectively. 

\item The radial distributions of median ICM density, velocity relative to
  the ICM and RP in the gas-particles method follow smooth profiles at
  $z$~=~0, and these distributions are very well traced by analytical
  estimations based on NFW profiles for the ICM density. However, the
  agreement is not so good for $z \geq$~1, where the shape of the RP
  profiles are similar but the mean values and levels of the percentiles
  shift to higher values in the analytical model. We find that the
  analytical calculation systematically overestimates the RP when compared
  to our self-consistent numerical approach. This overestimation grows
  with increasing redshift.

\item The distribution of RP values experienced by satellite galaxies
  within \rvir, estimated by both the gas-particles and the analytic
  methods, can be well fitted by Gaussian functions. The mean values of RP
  obtained from both methods are higher for the more massive clusters, a
  feature present for all redshifts. However, the difference
  between the two models becomes larger as one goes to higher redshifts, and
  with decreasing halo mass. This happens because the ICM density profiles
  in less massive clusters, and their progenitors at higher redshift, are
  galaxy-group sized systems of virial
  mass~$\sim$~10$^{13}-10^{14}\,h^{-1}$~\msun, whose ICM distributions are
  not so well described by the analytical profile adopted. In the
  analytical model the evolution of the mean RP with redshift is milder,
  especially in the case of the G14~clusters.

\item For the gas-particles method, the RP values at $z$~=~0 are weak
  (4.68$\times10^{-12}\,h^{2}$ dyn cm$^{-2}$) and medium
  (3.09$\times10^{-11}\,h^{2}$ dyn cm$^{-2}$) for G14 and G15~clusters,
  respectively. The majority of satellite galaxies in massive G15~clusters
  are already experiencing medium-level RP at $z \sim$~2, and at $z
  \lesssim$~0.5 about 20 per cent of satellites experience strong
  levels of RP (of order 10$^{-10}$~dyn~cm$^{-2}$). In the case of the
  G14~clusters, the mean RP is between medium and weak for $z
  \lesssim$~1.5, and few galaxies ever experience strong levels of RP;
  only at $z \sim$~1 does a significant fraction of satellites begin
  to experience medium levels of RP.

\item Based on our self-consistent approach, we find that RPS has a strong
  effect on the cold gas content of galaxies in both sets of clusters. At
  $z$~=~0 most galaxies ($\gtrsim$~70 per cent) within \rvir~are
  completely depleted of their cold star-forming gas. This is a strong
  difference with the model without RPS, where most galaxies manage to
  retain some cold gas: at the present epoch, only in the cluster
  cores ($r <$~0.5\rvir) the fractions of gas-depleted galaxies reach
  $\sim$~40 per cent. Observations of gas fractions as a function of
  clustercentric distance could provide strong constraints for the models.

\item The rate at which the cold gas is stripped from satellite galaxies
  depends on the virial mass of their host clusters. In our \sagrp~model,
  the fractions of gas-depleted galaxies for G14~clusters increase
  appreciably between $z$~=~1 and 0, whereas for G15~clusters the fractions
  at the present epoch are mostly established already at $z$~=~1.

\end{enumerate}

The general picture that emerges from the main results summarized above are
that the RPS effect depends on halo virial mass and redshift, being more
important in more massive haloes. Less massive galaxies within larger haloes
are the most affected, so this could be the mechanism responsible for the
transformation of dIrr galaxies into dSph in galaxy clusters
\citep{boselli2008}. In the more massive clusters the gas removal is
extremely effective. RPS could contribute to the pre-processing of galaxies
in smaller groups, before they fall into larger, cluster-sized
systems.

\section*{Acknowledgments}
The authors thank Klaus Dolag for making the simulations available to
us and for his help with this project, and the referee for useful
suggestions and comments that improved this paper. TET~wishes
to thank Ra\'ul Angulo, Carlton Baugh, Nelson Padilla and Leonardo
Pellizza for their useful comments and discussions. SAC~acknowledges
Simon White for originally proposing the ideas that led to this
project. This work was partially supported by PICT 245 Max Planck
(2006) and PROALAR 2007.

\bibliography{mn-jour,tecce_rps1_v1}

\label{lastpage}

\end{document}